\documentclass[letterpaper,draftclsnofoot,onecolumn,11pt]{IEEEtran}
\usepackage{amssymb, amsmath, epsfig, cite, epstopdf}

\newtheorem{lemma}{Lemma}
\newtheorem{theorem}{Theorem}
\newtheorem{definition}{Definition}

\newcommand  {\cC}  {{\cal C}}
\newcommand  {\cCccc} {{{\cal C}_\mathrm{ccc}}}
\newcommand  {\ccc} {{\mathrm{ccc}}}
\newcommand  {\avg} {{\mathrm{avg}}}
\newcommand  {\erf} {{\mathrm{erf}}}
\newcommand  {\cX}  {{\cal X}}
\newcommand  {\cY}  {{\cal Y}}
\newcommand  {\cP}  {{\cal P}}
\newcommand  {\cW}  {{\cal W}}
\newcommand  {\setS} {{\mathcal S}}
\newcommand  {\setH} {{\mathcal H}}
\newcommand  {\bc}  {{{\boldsymbol c}}}
\newcommand  {\br}  {{{\boldsymbol r}}}
\newcommand  {\wt}  {{\mathrm{wt}}}
\newcommand  {\ta} {{\mathrm{a}}}
\newcommand  {\td} {{\mathrm{d}}}
\newcommand  {\te} {{\mathrm{e}}}

\begin{document}

\IEEEoverridecommandlockouts 

\title{On the Throughput of Channels that Wear Out\thanks{This work is supported in part by the National Science Foundation under grant CIF-1623821.
Part of the work was presented at the 2017 IEEE International Symposium on Information Theory (ISIT) \cite{WuVT2017}.}
}

\author{Ting-Yi~Wu,
 	Lav~R.~Varshney,~\IEEEmembership{Senior Member,~IEEE}, and \\
	Vincent Y.~F.~Tan,~\IEEEmembership{Senior Member,~IEEE}
\thanks{Ting-Yi\ Wu is with the School of Electronics and Communication Engineering, Sun Yat-sen University, Guangzhou, China (e-mail: wutingyi@mail.sysu.edu.cn).
Lav R.\ Varshney is with the Department of Electrical and Computer Engineering and the Coordinated Science Laboratory, University of Illinois at Urbana-Champaign, Urbana, IL 61801, USA (e-mail: varshney@illinois.edu).
Vincent Y.~F.~Tan is with the Department of Electrical and Computer Engineering and Department of Mathematics, National University of Singapore, Singapore 117583
(e-mail: vtan@nus.edu.sg).
}
}
\maketitle
		 
\begin{abstract}
This work investigates the fundamental limits of communication over a noisy discrete memoryless channel that wears out, in the sense
of signal-dependent catastrophic failure.  In particular, we consider a channel that starts as a memoryless binary-input
channel and when the number of transmitted ones causes a sufficient amount of damage, the channel ceases
to convey signals. Constant composition codes are adopted to obtain an achievability bound and 
the left-concave right-convex inequality is then refined to obtain a converse bound on the  log-volume throughput for channels that wear out.
Since infinite blocklength codes will always wear out the channel for any finite threshold of failure and therefore cannot convey information at positive rates, 
we analyze the performance of finite blocklength codes to determine the maximum expected transmission volume 
at a given level of average error probability.
We show that this maximization problem has a recursive form 
and can be solved by dynamic programming. 
Numerical results demonstrate that a sequence of block codes is preferred to a single block code
for streaming sources.
\end{abstract}

%%%%%%%%%%%%%%%%%%%%%%%%%
\section{Introduction}
In reliability theory, there are two basic modes of catastrophic failure: \emph{independent} damage and \emph{cumulative} damage \cite{Nakagawa2007}.
With an independent damage process, a shock is either large enough to cause failure or it has no effect on the state of the system.  With a cumulative
damage process, however, each shock degrades the state of the system in an additive manner such that once the cumulative effect of all shocks
exceeds a threshold, the system fails.  Translating these notions to communication channels, failure can either be \emph{signal-independent} or \emph{signal-dependent}.  A typical channel with signal-dependent failure is in visible light communication under on-off signaling where light sources may burn out with `on' signals \cite{NairD2015}.

Here we consider optimizing communication over noisy channels that may wear out, i.e.\ suffer from signal-dependent failure.  As depicted in 
Table \ref{tab:classes}, this is a novel setting that is distinct from channels that die \cite{VarshneyMG2012} since failure time
is dependent on the realized signaling scheme, and from meteor-burst channels \cite{PursleyS1989,Ryan1997} and channels that heat up \cite{KochLS2009,OzelUG2016} since the channel noise level does not change with time.  The model is also distinct from Gallager's ``panic button'' \cite[p.~103]{Gallager1968} or ``child's toy'' \cite[p.~26]{Gallager1972} channel, since there is not a special input symbol that causes channel failure.

For example, consider a channel with finite input alphabet $\cX = \{0,1\}$ and finite output 
alphabet $\cY = \{0,1,?\}$.  It has an \emph{alive} state $\sigma = \ta$ when it acts like a 
binary symmetric channel (BSC) with crossover
probability $0 < \varepsilon < 1$, i.e.\ the transmission matrix is
\begin{equation}
\label{eq:pa}
p(y|x,\sigma=\ta) = p_\ta(y|x) = \begin{bmatrix} 1 - \varepsilon & \varepsilon & 0 \\
				\varepsilon & 1 - \varepsilon & 0 \end{bmatrix} \mbox{,}
\end{equation}
and a \emph{dead} state $\sigma = \td$ when it erases the input, i.e.\
the transmission matrix is
\begin{equation}
\label{eq:pd}
p(y|x,\sigma=\td) = p_\td(y|x) = \begin{bmatrix} 0 & 0 & 1 \\
				0 & 0 & 1 \end{bmatrix} \mbox{.}
\end{equation}

The channel starts in state $\sigma=\ta$ and then transitions to $\sigma=\td$ at some random time $T$,
where it remains for all time thereafter.  That is, the channel is in state $\ta$ 
for times $i = 1,2,\ldots,T$ and in state $\td$ for times $i = T+1,T+2,\ldots$. 
This failure time does not have a fixed and exogenous distribution $p_T(t)$, but depends on 
how the channel is used.  That is, the failure depends on the properties of the codeword that is transmitted
through the channel. 
When a $0 \in \cX$ is transmitted through the channel, the channel does not wear out 
whereas when a $1\in \cX$ is transmitted through the channel, the channel has a 
certain probability of getting damaged and moving closer to failure, as we detail in the sequel.  
\begin{table}
\centering
\caption{Classes of Channel Models}
\begin{tabular}{|c|c|c|}
\hline
 & {\bf signal-independent} & {\bf signal-dependent} \\ \hline
{\bf signal-to-noise} & meteor-burst channels\cite{PursleyS1989,Ryan1997} & channels that heat up\cite{KochLS2009,OzelUG2016} \\ \hline
{\bf failure time} & channels that die\cite{VarshneyMG2012} & channels that wear out \cite{WuVT2017}\\ \hline
\end{tabular}
\label{tab:classes}
\end{table}

Since it is inevitable for the channel to fail at a finite time
for any non-trivial signaling scheme, the Shannon capacity of the channel is 
zero.  Rather than invoking infinite blocklength asymptotic results, we must construct 
schemes that convey information via finite blocklength code(s)  
before the channel wears out.  Thus results in the finite blocklength regime 
\cite{Weiss1960,Strassen1962,PolyanskiyPV2010}
and their refinements \cite{TomamichelT2013, Moulin2017} 
can be built upon to determine limits on expected transmission
volume at a given average error probability.

Standard finite blocklength analysis, however,
cannot be directly applied since there is a restriction on transmitting too many $1$ symbols so that the channel stays alive. 
A principle of finite blocklength code design is therefore maximizing transmission volume while having a minimal number of $1$s.
To facilitate this, cost-constrained version of finite blocklength problems \cite{Fano1961,PolyanskiyPV2010,KostinaV2015,Moulin2012,Moulin2017} are studied,
probabilities of successfully transmitting a sequence of input-constraint codes
are also studied, and approximations of the fundamental communication limit of using constant composition codes 
in \cite{Moulin2012,ScarlettMF2015} and codes with input constraints \cite{KostinaV2015} are applied. 

To maximize the expected transmission volume, 
all possible sequences of finite-length constant composition codes with different input constraints 
have to be tested exhaustively.
Here we propose a recursive formulation to maximize the expected transmission volume in an efficient manner. 
The corresponding dynamic program 
and its graphical representation are provided. 
In considering the possibility of damage count feedback being available at the transmitter, 
we find that this does not change the probability of successfully transmitting a sequence of finite-length input-constraint codes
over the channel. Some numerical results are also provided to 
provide insights into the code design. 
From the numerical results, 
we observe that the sequence of codes which maximizes the expected volume follows the following rules: 
\begin{enumerate}
\item The code transmitted later has a shorter length,
\item The code transmitted later has a lighter Hamming weight. 
\end{enumerate}
These two observations are as intuitive, but not yet proven. 

The rest of this paper is organized as follows. 
Section \ref{sec:model} defines the problem statement including a specific focus on the wearing-out process.
The maximum expected transmission volume of using constant composition codes, 
which is treated as the achievable coding scheme, is discussed in Section \ref{sec:ccc}.
A dynamic programming formulation and discussion of feedback are given as well. 
Section \ref{sec:converse} investigates the converse bound, which removes the constraint of using constant composition codes. 
Section \ref{sec:num} provides some numerical results,
and Section \ref{sec:con} concludes this paper by suggesting some possible future directions. 

%%%%%%%%%%%%%%%%%%%%%%%%%%%%%%
\section{Channel Failure Model}
\label{sec:model}
Consider a channel with binary input alphabet $\cX=\{0,1\}$ and finite output alphabet $\cY=\{0,1,\dots,|\cY|-2, ?\}$,
and alive/dead states as indicated above.
There is a probability $\gamma$ of the channel getting damaged when a $1$ is transmitted through the channel. 
The channel starts at state $\sigma=\ta$ and transitions to state $\sigma=\td$
when the extent of damage exceeds a certain threshold $S$, 
where $S$ could be deterministic or random.
For simplicity, we regard $S$ as deterministic throughout this paper. 
The damage while transmitting a $1$ can be modeled by 
an independent Bernoulli random variable $D_k$ 
which takes the value $1$ with probability $\gamma$
and $0$ with probability $1-\gamma$.
Thus, the channel that wears out can be defined as a sextuple 
$(\cX, p_\ta, p_\td, \gamma, S, \cY)$.

The communication system over the channel that wears out $(\cX, p_\ta, p_\td, \gamma, S, \cY)$ is defined as follows. 
\begin{itemize}
\item An information stream is designed to be transmitted and it can be grouped into a sequence of $m$ messages,
$\left(W^{(1)}, W^{(2)}, \ldots, W^{(m)}\right)$. 
Each $W^{(i)}$ is chosen from the set $\cW^{(i)}=\{1,2,\ldots,M^{(i)}\}$ 
and transformed into the codeword $\bc^{(i)}$ with $n^{(i)}$ symbols by the encoder $f^{(i)}$, i.e.\
$f^{(i)}\left(W^{(i)}\right)=\bc^{(i)}\in\cX^{n^{(i)}}$.
The sequence of codewords, $\left(\bc^{(i)}\right)_{i=1}^m$ with $\sum_{i=1}^{m}n^{(i)}$ symbols in total, 
is then transmitted through the channel that wears out. 
Let such a sequence of $m$ codebooks be an $\left(M^{(i)}, n^{(i)}\right)_{i=1}^m$-code. 

\item The received sequence $\br\in\cY^{\sum_{i=1}^{m}n^{(i)}}$ is decoded into $\left(\hat{W}^{(1)}, \ldots, \hat{W}^{(m)}\right)$
by the decoders $g^{(i)}$, where \ $g^{(i)}\left(\br^{(i)}\right)=\hat{W}^{(i)}$ 
for $\br^{(i)}=\br_{n^{(1)}+\cdots +n^{(i-1)}+1}^{n^{(1)}+\cdots +n^{(i-1)}+n^{(i)}}$ and $1\leq i\leq m$. 
If all $n^{(i)}$ channel outputs for decoder $g^{(i)}$ are not $?$, then $\hat{W}^{(i)}\in\cW^{(i)}$;
otherwise,  $\hat{W}^{(i)}=\te$.
\item The average decoding error probability for a codebook $\left(M^{(i)}, n^{(i)}\right)$ is defined as 
\begin{equation}
P_\te^{(i)}=\frac{1}{M^{(i)}}\sum_{w\in\cW^{(i)}}\Pr\left[\hat{W}^{(i)}\neq w \middle| W^{(i)} = w \wedge \hat{W}^{(i)}\neq \te\right],
\end{equation}
and the decoding error probability for an $\left(M^{(i)}, n^{(i)}\right)_{i=1}^m$-code is defined as 
\begin{equation}
P_\te=\max_{i\in\{1,\ldots, m\}} P_\te^{(i)}.
\end{equation}
An $\left(M^{(i)}, n^{(i)}\right)_{i=1}^m$-code is said to be $\eta$-achievable if $P_\te\leq\eta$.
\end{itemize}

Let the sequence of codewords $\left(\bc^{(i)}\right)_{i=1}^m$ be chosen from an $\eta$-achievable $\left(M^{(i)}, n^{(i)}\right)_{i=1}^m$-code,
then the channel wears out when  
\begin{equation}
\sum_{i=1}^{m}\sum_{k=1}^{\wt \left( \bc^{(i)} \right)}D_k> S,
\end{equation}
where $\wt \left( \bc^{(i)} \right)$ is the Hamming weight of codeword $\bc^{(i)}$.

Let $B(S, h, \gamma)$ be the probability of the channel staying alive after $h$ ones are transmitted. That is,
\begin{equation}
B(S,h,\gamma)\triangleq\Pr\left[\sum_{k=1}^{h}D_k\leq S\right].
\end{equation}
$B(S,h,\gamma)$ is the cumulative distribution function of the binomial distribution 
if $S$ is deterministic. 
Hence, for channel that wears out $(\cX, p_\ta, p_\td, \gamma, S, \cY)$,
the probability of not wearing out the channel after transmitting the sequence of codewords $\left(\bc^{(i)}\right)_{i=1}^m$ is
\begin{IEEEeqnarray}{rCl}
\Pr\left[\left(\bc^{(i)}\right)_{i=1}^{m}\mbox{ alive}\right] &\triangleq&
\Pr\left[ \sum_{i=1}^{m}\sum_{k=1}^{\wt\left(\bc^{(i)}\right)}D_k\leq S \right]\\
&=&B\left(S,\sum_{i=1}^{m}\wt\left(\bc^{(i)}\right),\gamma\right).
\end{IEEEeqnarray}
Let the $\eta$-achievable $\left(M^{(i)}, n^{(i)}\right)_{i=1}^m$-code be denoted as $\left(\cC^{(i)}\right)_{i=1}^m$ and
all the codewords in the same $\cC^{(i)}$ are transmitted with equal probability,
the average probability of successfully transmitting $\left(\cC^{(i)}\right)_{i=1}^m$ is
\begin{equation}\label{eqn:alivepro}
\Pr\left[\left(\cC^{(i)}\right)_{i=1}^{m}\mbox{ alive}\right] \triangleq \left\{\prod_{i=1}^m \frac{1}{M^{(i)}}\right\}\times
\left\{\sum_{\left(\bc^{(i)}\right)_{i=1}^{m}\in \left(\cC^{(i)}\right)_{i=1}^{m}}
B\left(S,\sum_{i=1}^{m}\wt\left(\bc^{(i)}\right),\gamma\right)\right\}.
\end{equation}
Based on the results in \cite{VarshneyMG2012}, 
the expected transmission log-volume $V$ of transmitting $\left(\cC^{(i)}\right)_{i=1}^m$ 
over the channel that wears out $(\cX, p_\ta, p_\td, \gamma, S, \cY)$ 
at a given level of error probability $\eta$ can be derived as
\begin{equation}\label{eqn:vol}
V\left(\left(\cC^{(i)}\right)_{i=1}^{m}\right) = \sum_{j=1}^{m}\Pr\left[\left(\cC^{(i)}\right)_{i=1}^{j}\mbox{ alive}\right]\log M^{(j)}.
\end{equation}

%%%%%%%%%%%%%%%%%%%%%%%%%%%%
\section{Achievability Bound}
\label{sec:ccc}
To obtain an achievability bound on $V$ in \eqref{eqn:vol}, we restrict attention to finite blocklength {\it constant composition codes}, denoted as $\cCccc$,
in which all codewords from the same codebook have the same number of ones.
Given a $\cCccc$ with length $n$, 
the Hamming weight of each codeword can be denoted as $\wt(P)\triangleq nP(1)$,
where $P$ is a type from $\cP_{n}(\cX)$, the set of all types formed from sequences of length $n$.
Define an $(n,M,\eta)_P$-code to be an $\eta$-achievable constant composition code with type $P$, 
blocklength $n$, number of messages $M$, and average error probability no larger than $\eta$.
Hence, when the constant composition code corresponding to $P$ is transmitted, 
the damage count for such a code is 
\begin{equation}
U(P)\triangleq\sum_{k=1}^{\wt(P)}D_k. 
\end{equation}
Suppose an $\left((n^{(i)},M^{(i)},\eta)_{P^{(i)}}\right)_{i=1}^m$-code, denoted as $\left(\cC_\ccc^{(i)}\right)_{i=1}^m$,
is conveyed through the channel.
The individual codes need not be the same, and so the full concatenation is much like a constant subblock composition code \cite{TandonMV2016}. 
The probability of the channel $(\cX, p_\ta, p_\td, \gamma, S, \cY)$ staying alive 
after conveying the first $j$ codes $\left(\cC^{(i)}_\ccc\right)_{i=1}^j$ in \eqref{eqn:alivepro}
can be further written as
\begin{equation}\label{eqn:aliveproccc}
\Pr\left[\left(\cC^{(i)}_\ccc\right)_{i=1}^{j}\mbox{ alive}\right]=B\left(S,\sum_{i=1}^{j}\wt\left(P^{(i)}\right),\gamma\right).
\end{equation}
Similar to the result in \cite{VarshneyMG2012}, 
the expected log-volume for transmitting $\left(\cC^{(i)}_\ccc\right)_{i=1}^m$ 
with a maximum average error probability $\eta$ can be expressed as
\begin{multline}\label{eqn:volccc}
\sum_{j=1}^{m}\Pr\left[\left(\cC^{(i)}_\ccc\right)_{i=1}^{j}\mbox{ alive}\right]\log M_\ccc^{*}\left(n^{(j)}, \wt\left(P^{(j)}\right),\eta\right)\\=
\sum_{j=1}^{m}B\left(S,\sum_{i=1}^{j}\wt\left(P^{(i)}\right),\gamma\right)\log M_\ccc^{*}\left(n^{(j)}, \wt\left(P^{(j)}\right),\eta\right)
\end{multline}
where 
\begin{equation}
 M_\ccc^{*}(n, \wt(P),\eta)=\max\left\{M |\exists \mbox{ an } (n, M, \eta)_P\mbox{-code} 
 \mbox{ for the alive channel } p_\ta \right\}
\end{equation} 
is the maximum transmission volume of the constant composition code over the binary-input DMC
when the channel is alive.

To maximize the expected log-volume in \eqref{eqn:volccc} given a total length $N=\sum_{i=1}^m n^{(i)}$, 
a $\left(\cC^{(i)}_\ccc\right)_{i=1}^m$ for $1\leq m \leq N$ to maximize \eqref{eqn:volccc} needs to be found. 
Let $Z_\ccc(N,H,\eta)$ be the set of all possible $\left((n^{(i)},M^{(i)},\eta)_{P^{(i)}}\right)_{i=1}^m$-codes 
for all $m\in\{1,\ldots,N\}$, 
which have total length $N$ and total Hamming weight $H$, i.e.,
\begin{multline}\label{eqn:zccc}
Z_\ccc(N, H,\eta)\triangleq \left. \Bigg\{ \left(\cC^{(i)}\right)_{i=1}^{m} \middle| \right.
 \cC^{(i)} \mbox{ is an } \left(n^{(i)}, M^{(i)}, \eta\right)_{P^{(i)}}\mbox{-code}, 0< m < N, \\
\left. \sum_{i=1}^{m} n^{(i)}=N  \mbox{ and } \sum_{i=1}^{m}\wt\left(P^{(i)}\right)=H\Bigg\}\right..
\end{multline}
Then the maximum expected log-volume with the given $N$ and $H$ 
of transmitting the constant composition codes is denoted as follows:
\begin{equation}
\label{eqn:volccc_}
V_\ccc^*(N,H, \eta)=
\max_{\substack{\left(\cC^{(i)}\right)_{i=1}^{m}\in\\ Z_\ccc(N,H,\eta)}} 
\left\{\sum_{j=1}^{m}B\left(S,\sum_{i=1}^{j}\wt\left(P^{(i)}\right),\gamma\right)
\log M_\ccc^{*}\left(n^{(j)}, \wt\left({P^{(j)}}\right),\eta\right)\right\},
\end{equation}
and the maximum expected log-volume with $N$ is 
\begin{equation}\label{eqn:volccc_V}
V^*_\ccc(N, \eta)=\max_{0< H<N} V_\ccc^*(N,H, \eta).
\end{equation}
It should be noted that $m$ does not need to be specified explicitly in the maximization in \eqref{eqn:volccc_} since $Z_\ccc(N, H, \eta)$
in \eqref{eqn:zccc} consists of all collections of codes $\left(C_{\ccc}^{(i)}\right)_{i=1}^m$ for each $1\le m \le N-1$.

\subsection{Dynamic Programming Formulation}\label{sec:dpccc}
To solve the maximization problem in equation \eqref{eqn:volccc_}, a dynamic programming formalism is adopted. 
A recursive form of \eqref{eqn:volccc_} can be formulated as: 
\begin{equation}
\label{eqn:dprec}
V^*_\ccc(N,H,\eta)=\max_{\substack{\text1\leq n\leq N\\1\leq h \leq H}}\left\{V^*_\ccc(N-n, H-h,\eta)+
B\left(S,H,\gamma\right) \log M_\ccc^*(n,\wt(P),\eta)\right\},
\end{equation}
where $P(1)=h/n$.

\begin{figure}[t]
    \centering
    \includegraphics[width=4.5in]{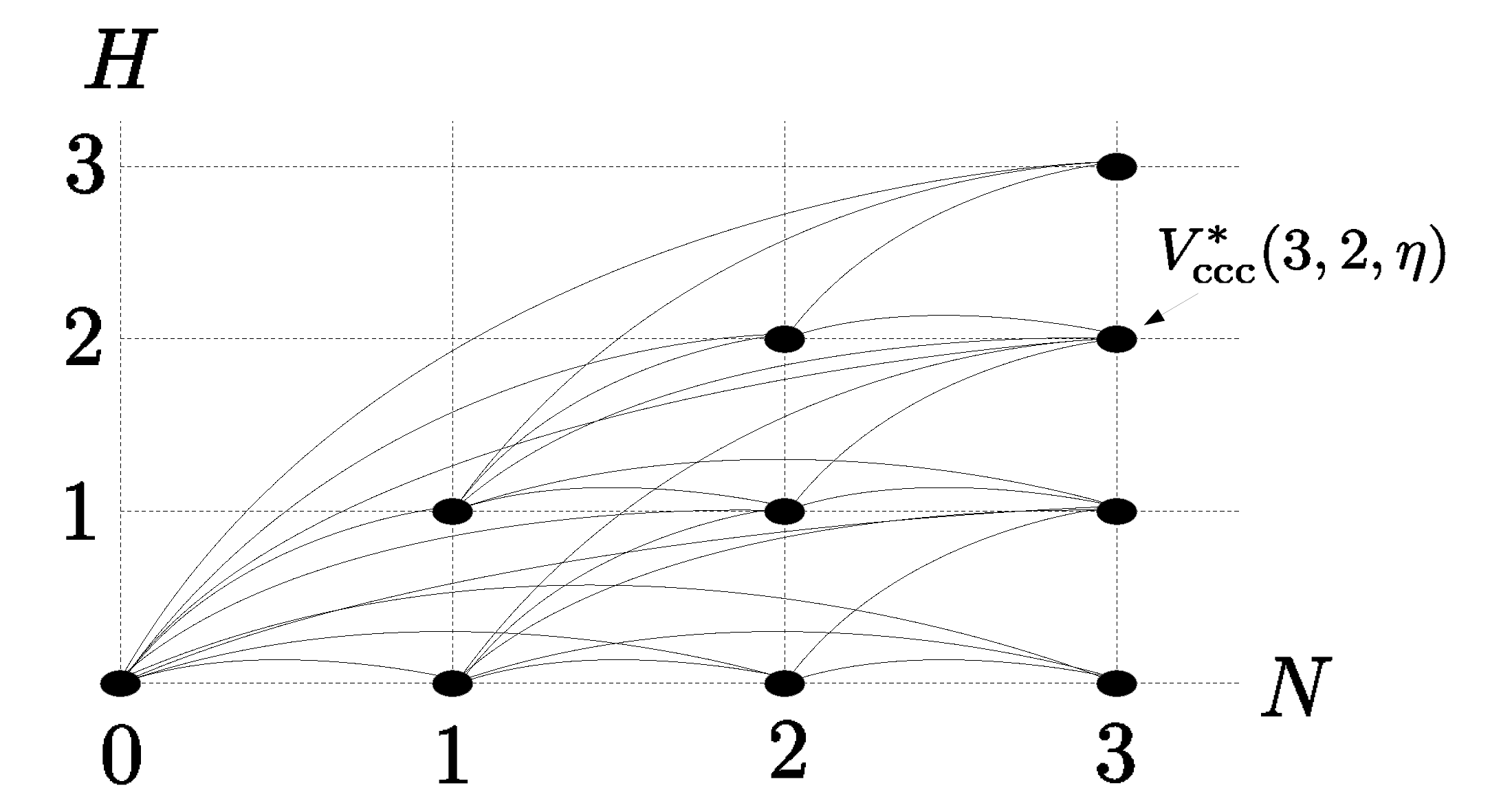}
    \caption{Graphical representation of the recursive form \eqref{eqn:dprec}.}
    \label{fig:dptrellis}
\end{figure}

A graphical representation of the recursive form \eqref{eqn:dprec} is illustrated in Fig.~\ref{fig:dptrellis}.
In this trellis diagram, the metric of the branch from node $(x_1, y_1)$ to node $(x_2, y_2)$ is 
\begin{equation}
B\left(S,y_2,\gamma\right) \log M_\ccc^*(x_2-x_1,\wt(P), \eta),
\end{equation}
where $P(1)=\frac{y_2-y_1}{x_2-x_1}$.

The path with the maximum accumulated branch metric from node ($0,0$) to node ($N,H$) is the solution for $V^*_\ccc(N,H,\eta)$. 
Thus the optimization problem in \eqref{eqn:volccc_} can be reduced to finding the longest path in Fig.~\ref{fig:dptrellis}, 
in which a dynamic programming algorithm based on \eqref{eqn:dprec} is applied to break down the problem $V^*_\ccc(N,H,\eta)$ 
into easier subproblems $V^*_\ccc(N-n, H-h,\eta)$ recursively. 

\subsection{Damage Count Feedback Does Not Improve Performance}
The expected log-volume in \eqref{eqn:volccc} contains two parts: 
the probability of successfully conveying $\cC^{(i)}$ and the volume of $\cC^{(i)}$.
How the volume of $\cC^{(i)}$ over DMCs is affected 
by full feedback has been studied in \cite{AltugW2014}, \cite[Ch.~20]{PolyanskiyW}, 
which is beyond the scope of this paper.
Instead, the following structural question regarding damage count feedback (rather than full feedback) is raised: 
{\it Will the probability of successfully conveying $\cC^{(i)}$ be different 
due to feedback of the damage count?}
 
Suppose $\cC^{(1)}$ has been sent through the channel 
and instant feedback tells the transmitter the damage count $d$ caused by transmitting $\cC^{(1)}$ 
before conveying $\cC^{(2)}$.
The probability of causing $d$ damages from $\wt\left( P^{(1)}\right)$ transmitted $1$s is
\begin{equation}\label{eqn:profeedback}
\Pr\left[ U(P^{(1)}) = d\right].
\end{equation}
Since the transmitter knows that $d$ damage events have happened,
the channel is still capable of handling $S-d$ damages. 
The probability of successfully transmitting $\cC^{(2)}$ given $d$ damages without wearing out the channel is
\begin{equation}\label{eqn:protranssec}
\Pr\left[ U(P^{(2)})\leq S-d \right].
\end{equation}
Combining \eqref{eqn:profeedback} and \eqref{eqn:protranssec}, 
the overall probability of successfully transmitting $\cC^{(2)}$ is
\begin{IEEEeqnarray}{rCl}
&&\sum_{d=0}^{S} \Pr\left[ U(P^{(1)}) = d\right]\times\Pr\left[ U(P^{(2)})\leq S-d \right]\\
&&=\Pr\left[ U(P^{(1)})+U(P^{(2)}) \leq S \right]\\
&&=B\left(S,\wt(P^{(1)})+\wt(P^{(2)}),\gamma\right).
\end{IEEEeqnarray}
The above result can be extended to have the following probability of successfully conveying $\cC^{(i)}$
\begin{equation}
B\left(S,\sum_{j=1}^{i}\wt\left(P^{(j)}\right),\gamma\right),
\end{equation}
which coincides with \eqref{eqn:aliveproccc}.
Hence the probability of successfully transmitting $\cC^{(i)}$ remains the same with or without damage state feedback. 

This implies that the achievable transmission volume cannot be increased by providing damage state feedback.

%%%%%%%%%%%%%%%%
\section{Converse Bound}\label{sec:converse}
To obtain a converse (upper) bound on the log-volume for a channel that wears out, 
we bound the alive probability and the transmission volume in \eqref{eqn:vol} separately from above. 
Without the constraint of using constant composition codes, 
the Hamming weights of codewords may be different from each other.
Since the transmission volume $M^{(j)}$ in \eqref{eqn:vol} with a given Hamming weight spectrum 
cannot be obtained in closed-form, we turn our attention to codes with a given average Hamming weight constraint.

%%%
\subsection{Upper Bound on Transmission Volume}
Let $(n, M, \eta)$-codes be $\eta$-achievable codes of length $n$ and size $M$.
Denote the Hamming weight of the $i^{\mathrm{th}}$ codeword by $w_i$ for all $i\in\{1,2,\ldots,M\}$.  
The maximum size of $(n, M, \eta)$-codes with average Hamming weights not exceeding $w^*\triangleq\sum_{i=1}^M w_i$ is given as
\begin{multline}
M_{\avg}^*(n, w^*, \eta)=\max\{M |\exists \mbox{ an } (n, M, \eta)\mbox{-code}
 \mbox{ for the alive channel } p_\ta \\\mbox{ such that its average Hamming weight $\leq w^*$}\}.
\end{multline}
The normal approximation of $M_{\avg}^*(n, w^*, \eta)$ can be evaluated as \cite[Eq.~(1)]{KostinaV2015}.
Clearly, $M_{\avg}^*(n, w^*, \eta)$ is a non-decreasing function of $w^*$.

%%%%%%%%%%%%%%%%%%%%%%%%%%%
\subsection{Upper Bound on Alive Probability}
Now we find an upper bound for the alive probability \eqref{eqn:alivepro}
with a given average Hamming weight constraint.
Given a sequence of codes $\big(\cC^{(i)}\big)_{i=1}^{m}$ with the average Hamming weight $w^*_m$ defined as
\begin{equation}\label{eqn:averageweight}
w^*_m=\left\{\prod_{i=1}^m \frac{1}{M^{(i)}}\right\}\times
\left\{\sum_{\left(\bc^{(i)}\right)_{i=1}^{m}\in \left(\cC^{(i)}\right)_{i=1}^{m}}
\sum_{i=1}^{m}\wt\left(\bc^{(i)}\right)\right\},
\end{equation}
our objective is to upper bound \eqref{eqn:alivepro}. 
Since the binomial cdf $B(S, w, \gamma)$ is discrete in $w$,
an upper bound is difficult to obtain.
Hence we use the Berry-Esseen inequality to obtain an upper bound for the binomial cdf \cite[Theorem 1]{Schulz2016}. 
This inequality states that 
\begin{equation}\label{berryessen}
B(S, w, \gamma)\leq B_\mathrm{N}(S,w,\gamma)+B_\mathrm{BE}(w),
\end{equation}
where 
\begin{IEEEeqnarray}{rCl}
B_\mathrm{N}(S,w,\gamma)&=&\frac{1}{2}\Bigg[1+\erf\Bigg(\frac{S-w\gamma}{\sqrt{2w\gamma(1-\gamma)}}\Bigg)\Bigg],\\
B_\mathrm{BE}(w)&=&\frac{\sqrt{10}+3}{6\sqrt{2\pi}}\cdot\frac{\gamma^2+(1-\gamma)^2}{\sqrt{w\gamma(1-\gamma)}},
\end{IEEEeqnarray} 
and $\erf (\cdot)$ is the Gaussian error function. 
However, both $B_\mathrm{N}(S,w,\gamma)$ and $B_\mathrm{BE}(w)$ are undefined at $w=0$ 
and $B_\mathrm{BE}(w)$ is greater than $1$ when $w$ is small. 
To rule out the undefined point and tighten the bound, we replace $B_\mathrm{N}(S,w,\gamma)$ and $B_\mathrm{BE}(w)$ respectively by
\begin{equation}\label{Bbar_N}
f_\mathrm{N}(w)=
\begin{cases}1 & \text{if } w=0,\\ 
B_\mathrm{N}(S,w,\gamma) & \text{otherwise},
\end{cases}
\end{equation}
and
\begin{equation}\label{Bbar_BE}
f_\mathrm{BE}(w)=
\begin{cases}1 & \text{if } 0\leq w\leq w_\mathrm{BE},\\ 
B_\mathrm{BE}(w) & \text{if }w_\mathrm{BE}<w,
\end{cases}
\end{equation}
where $w_\mathrm{BE}=\max\{w|B_\mathrm{BE}(w)\geq 1\}$, and the inequality 
$B(S, w, \gamma)\leq f_\mathrm{N}(w)+f_\mathrm{BE}(w)$ still holds. 

As shown in \eqref{eqn:alivepro}, the alive probability is calculated as the average of $B(S, w, \gamma)$.
To simplify the problem formulation, we consider upper bounding $\frac{1}{M}\sum_{i=1}^{M}B(S, w_i, \gamma)$
for a given $\eta$-achievable $(n, M, \eta)$-code with weight spectrum $\{w_i\}_{i=1}^M$.
Clearly $w_i\in\{0,1,\ldots,n\}$ for all $i$.
From \eqref{berryessen}, \eqref{Bbar_N}, and \eqref{Bbar_BE}, we obtain
\begin{IEEEeqnarray}{rCl}
\frac{1}{M}\sum_{i=1}^{M}B(S, w_i, \gamma)&\leq&\frac{1}{M}\sum_{i=1}^{M}f_\mathrm{N}(w_i)+f_\mathrm{BE}(w_i)\\
&=&\frac{1}{M}\sum_{i=1}^{M}f_\mathrm{N}(w_i)+\frac{1}{M}\sum_{i=1}^{M}f_\mathrm{BE}(w_i).\label{eqn:berryesseenbound}
\end{IEEEeqnarray}
Now we introduce left-concave right-convex (LCRC) functions \cite[Sec.~3.3]{Cirtoaje2006}.
\begin{definition}\label{def:LCRC}
$f:[a,\infty) \rightarrow \mathbb{R}$ is said to be an LCRC function if
it is continuous on $[a,\infty)$ and there exists a $c\in[a,\infty)$ such that 
$f$ is concave on $[a,c]$ and convex on $[c,\infty)$.
\end{definition}

From \eqref{Bbar_N} and \eqref{Bbar_BE}, it is clear that both $f_\mathrm{N}(w)$ and $f_\mathrm{BE}(w)$ are well-defined
for all nonnegative real values $w$. 
We extend the definitions of $f_\mathrm{N}$ and $f_\mathrm{BE}$ so that their domains are $[0,n]$
and write $f_\mathrm{N}(x)$ and $f_\mathrm{BE}(x)$ for $x\in[0,n]$.
Clearly $f_\mathrm{N}$ and $f_\mathrm{BE}$ are continuous.
The following two lemmas state that both $f_\mathrm{N}(x)$ and $f_\mathrm{BE}(x)$ are non-increasing LCRC functions of $x$.
\begin{lemma}\label{lm:LCRC-normalapprox}
The function $f_\mathrm{N}:[0,\infty) \rightarrow \mathbb{R}$ is a non-increasing LCRC function.
%%%%%%%%%
\end{lemma}
\begin{IEEEproof}
See Appendix \ref{append:LCRC-normalapprox}.
\end{IEEEproof}
%%%%%%%%%
\begin{lemma}\label{lm:LCRC-berryesseen}.
The function $f_\mathrm{BE}:[0,\infty) \rightarrow \mathbb{R}$ is a non-increasing LCRC function.
\end{lemma}
\begin{IEEEproof}
See Appendix \ref{append:LCRC-berryesseen}.
\end{IEEEproof}

Now let us recall Karamata's majorization inequality \cite{Karamata1932}. 
%%%%%%%%%
\begin{lemma}[Karamata's majorization inequality]
\label{lm:karamata}
Given $a_1\geq a_2\geq \cdots \geq a_M$ and $b_1\geq b_2\geq \cdots \geq b_M$ such that $a_i$, $b_i$ are in interval $I$.
Let $A_h\triangleq\sum_{i=1}^{h}a_i\geq B_h\triangleq\sum_{i=1}^{h}b_i$ for all $1\leq h\leq M-1$, 
and $A_M\triangleq\sum_{i=1}^{M}a_i=B_M\triangleq\sum_{i=1}^{M}b_i$,
then 
\begin{equation}
\sum_{i=1}^{M}f(a_i) \geq \sum_{i=1}^{M}f(b_i)
\end{equation}
if $f$ is a convex function on $I$.
\end{lemma}
\begin{IEEEproof}
See Appendix \ref{append:karamata} for a self-contained proof. 
\end{IEEEproof}

Given two tuples $(a_1,a_2,\ldots,a_M)$ and $(b_1,b_2,\ldots,b_M)$ such that 
$\sum_{i=1}^{M}a_i=\sum_{i=1}^{M}b_i$, we say that $(a_1,a_2,\ldots,a_M)$ majorizes $(b_1,b_2,\ldots,b_M)$ 
if $\sum_{i=1}^{h}a_i\geq \sum_{i=1}^{h}b_i$ for all $1\leq h\leq M-1$.
The following lemma specifies the sequence which majorizes all sequences with the same average.
%%%%%%%%%
%Majorization Sequence
%%%%%%%%%
\begin{lemma}\label{lm:majorsequence}
For $c,b\in\mathbb{R}$ such that $c<b$, any tuple in $[c,b]^M$ with average $x^*$ is majorized by the following sequence of length $M$.  
\begin{equation}\label{eqn:majorsequence}
x_i=
\begin{cases}
b & \text{if } 1\le i\le j,\\ 
Mx^*-jb-(M-j-1)c & \text{if } i= j+1,\\
c & \text{otherwise,} 
\end{cases}
\end{equation}
where $j=\left\lfloor \frac{M(x^*-c)}{b-c} \right\rfloor$.
\end{lemma}
\begin{IEEEproof}
See Appendix \ref{append:majorsequence}. 
\end{IEEEproof}

\begin{figure}[t]
    \centering
    \includegraphics[width=4.5in]{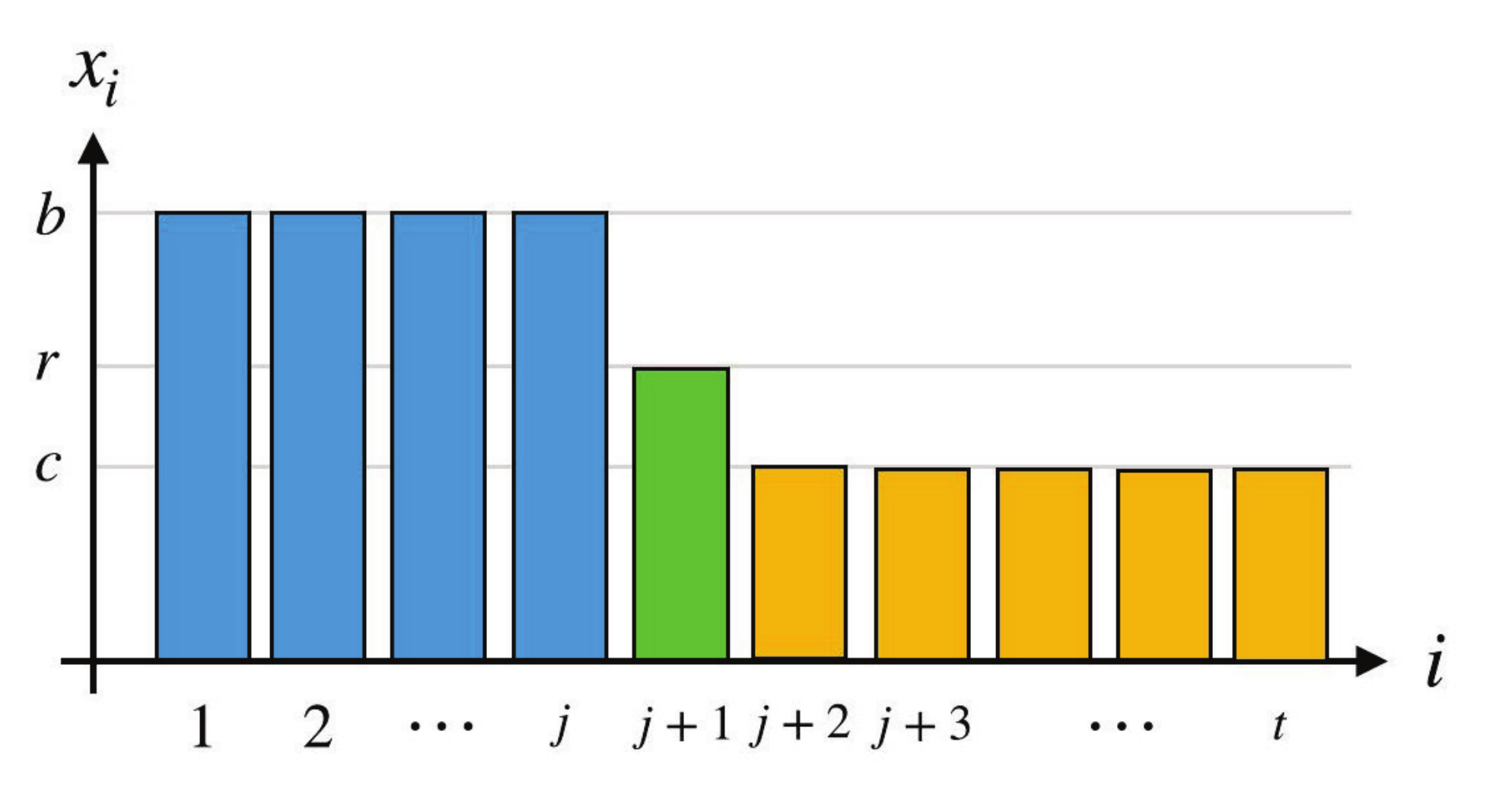}
    \caption{An illustration of the majorizing sequence \eqref{eqn:majorsequence}, which separates all $x_i$ into three categories. 
    $j$ blue bars, a green bar, and $t-j-1$ yellow bars denote the $x_i$ equal to $b$, $r=Mx^*-jb-(M-j-1)c$, and $c$, respectively.}
    \label{fig:majorizedseq}
\end{figure}

It can be verified that $c\le Mx^*-jb-(M-j-1)c \le b$. 
An illustration of the sequence \eqref{eqn:majorsequence} is given in Fig.~\ref{fig:majorizedseq}, 
in which $r=Mx^*-jb-(M-j-1)c$. 
Fig.~\ref{fig:majorizedseq} shows that the sequence majorizes other sequences with the same average has the property that it 
keeps assigning the largest value $b$ to $x_i$ from $i=1$ until the constraint $\sum_{i=1}^Mx_i =Mx^*$ is no longer satisfied.

\begin{figure}[t]
    \centering
    \includegraphics[width=6in]{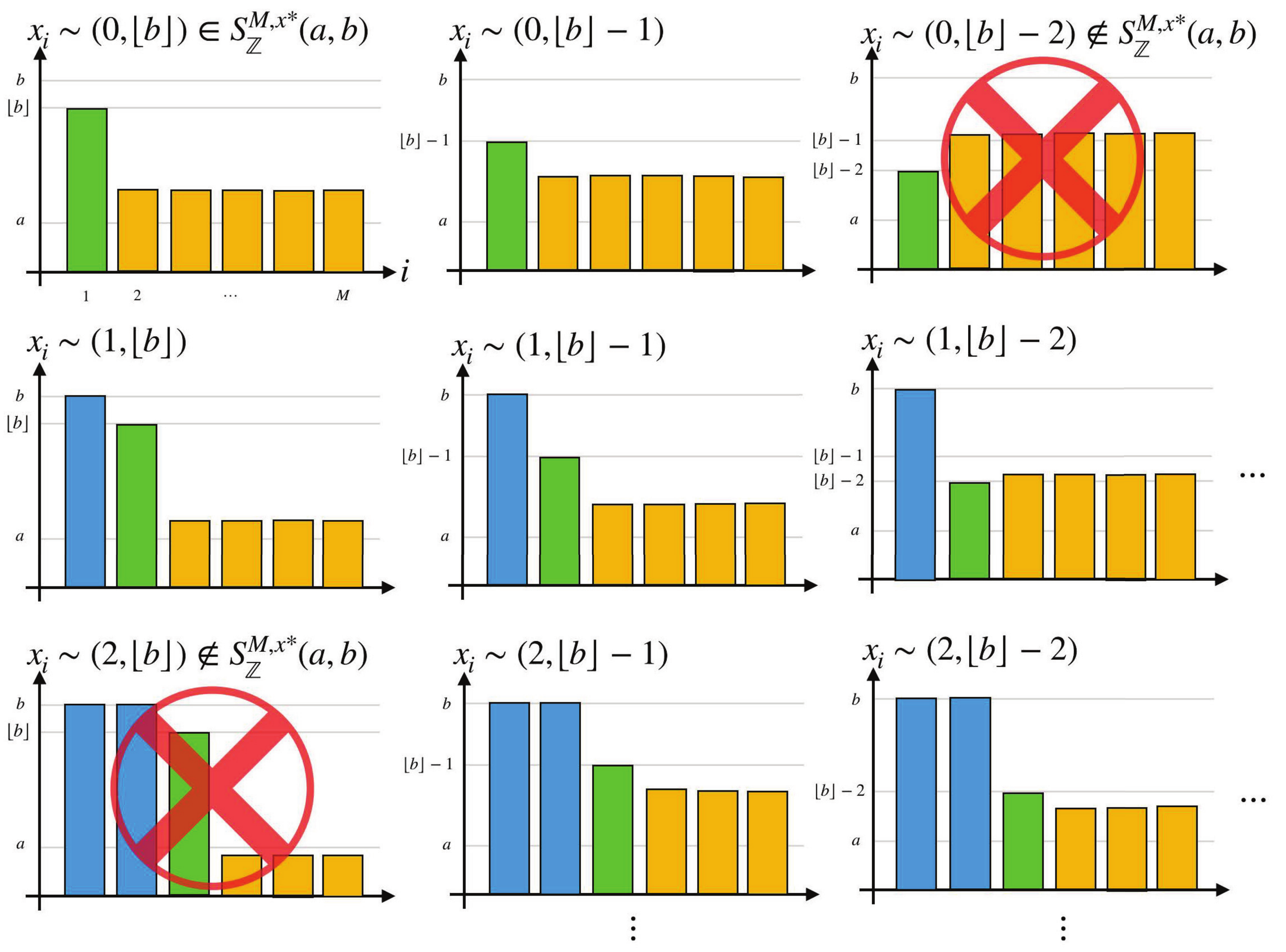}
    \caption{An illustration of the sequences corresponding to $(j,k)\in\setS^{M,x^*}_\mathbb{Z}(a,b)$. 
    All the sequences corresponding to the $(j,k)\notin\setS^{M,x^*}_\mathbb{Z}(a,b)$ are not considered in the maximization \eqref{eqn:col_lcrc_integer}.}
    \label{fig:lcrcseq}
\end{figure}

Combining Lemmas \ref{lm:karamata} and \ref{lm:majorsequence}, 
the sequence \eqref{eqn:majorsequence} has the maximal $\sum_{i=1}^Mf(x_i)$ among all sequences from $[a,b]^M$ if function $f$ is convex on $[a,b]$.
Given an LCRC function $f$ and the average $\frac{1}{M}\sum_{i=1}^M x_i$, 
Jensen's inequality and Karamata's inequality can be applied to upper bound
$\frac{1}{M}\sum_{i=1}^M f(x_i)$ as stated in \cite[Sec.~3.3]{Cirtoaje2006} within all $x_i\in[a,\infty)$ (LCRC inequality). 
However, according to \eqref{eqn:berryesseenbound}, all $x_i$ are restricted to be in the bounded interval $[0,n]$. 
To fit our purpose, we revise the LCRC inequality to upper bound $\sum_{i=1}^M f(x_i)$ 
where all $x_i$ are drawn from a bounded interval $[a,b]$. 
%%%%%%%%%%%%
%LCRC inequality
%%%%%%%%%%%%
\begin{lemma}[Revised LCRC inequality]
\label{lm:lcrc}
Given $a,b,c\in\mathbb{R}$ such that $a<b$ and $a<c$. 
Let $f$ be a continuous LCRC function on $[a,\infty)$ and $c$ be the point separating the concave region and the convex region.
Given $x_i\in [a,b]$ for $1\leq i \leq M$ and $\frac{1}{M}\sum_{i=1}^M x_i=x^*$, then
\begin{equation}\label{eqn:col_lcrc_real}
\sum_{i=1}^M f(x_i)\leq \max_{(j,r)\in\setS^{M,x^*}_\mathbb{R}(a,b)}\left\{jf(b)+f(r)+(M-j-1)f\left(\frac{Mx^*-jb-r}{M-j-1}\right)\right\},
\end{equation}
where 
\begin{equation}
\setS^{M,x^*}_\mathbb{R}(a,b)=\left\{(j,r)\,\middle|\, j\in\{0,1,\ldots,M-1\},\  r\in\mathbb{R} \mbox{ s.t. } b\ge r\ge\left(\frac{Mx^*-jb-r}{M-j-1}\right)\ge a\right\}. 
\end{equation}
If $f$ is also a non-increasing function and $a$ is an integer, inequality \eqref{eqn:col_lcrc_real} can be rewritten as
\begin{equation}\label{eqn:col_lcrc_integer}
\sum_{i=1}^M f(x_i)\leq \max_{(j,k)\in\setS^{M,x^*}_\mathbb{Z}(a,b)}\left\{jf(b)+f(k)+(M-j-1)f\left(z\right)\right\},
\end{equation}
where $z=\max\left\{\frac{Mx^*-jb-k-1}{M-j-1}, a\right\}$ and
\begin{equation}
\setS^{M,x^*}_\mathbb{Z}(a,b)=\left\{(j,k)\,\middle|\, j\in\{0,1,\ldots,M-1\},\  k\in\{a,a+1, \ldots,\lfloor b\rfloor \} \mbox{ s.t. } k+1\ge z\right\}. 
\end{equation}
\end{lemma}
\begin{IEEEproof}
See Appendix \ref{append:lcrc}.
\end{IEEEproof}
Given some $(j,k)$, the corresponding sequences $(x_1,x_2,\ldots,x_M)$ are depicted in Fig.~\ref{fig:lcrcseq}. 
As shown in Fig.~\ref{fig:lcrcseq}, two instances are of $(j,k)\notin\setS^{M,x^*}_\mathbb{Z}(a,b)$, they are $(0,\lfloor b\rfloor -2)$ and $(2,\lfloor b\rfloor)$.
For the instance of $(j,k)=(0,\lfloor b\rfloor -2)$, which is not in $\setS^{M,x^*}_\mathbb{Z}(a,b)$ because of $k+1< z $, 
hence the corresponding sequence can not maximize $\sum_{i=1}^M f(x_i)$. 
For the other instance of $(j,k)=(2,\lfloor b\rfloor)$, which is not in $\setS^{M,x^*}_\mathbb{Z}(a,b)$ due to $x_i\notin[a,b]$ for $i\ge j+2$. 
The maximization \eqref{eqn:col_lcrc_integer} only considers those sequences corresponding to valid $(j,k)$.

With the help of Lemma \ref{lm:lcrc}, we now can upper bound the alive probability in \eqref{eqn:alivepro} as follows.
\begin{theorem}\label{thm:aliveproconverse}
Given a sequence of codes, $\left(\cC^{(i)}\right)_{i=1}^{m}$ with average Hamming weights $w^*_m$,
the alive probability of conveying $\left(\cC^{(i)}\right)_{i=1}^{m}$ is upper-bounded as follows.
\begin{IEEEeqnarray}{rCl}
&\Pr&\left[\left(\cC^{(i)}\right)_{i=1}^{m}\mbox{ alive}\right] \nonumber\\
&\leq&\frac{1}{M} \max_{(j,k)\in\setS^{M,w_m^*}_{\mathbb{Z}}(0,N)}\left\{jf_{\mathrm{N}}(N)+f_{\mathrm{N}}(k)+(M-j-1)f_{\mathrm{N}}\left(\frac{Mw^*_m-jN-k-1}{M-j-1}\right)\right\}\nonumber\\
&&+\frac{1}{M}\max_{(j,k)\in\setS^{M,w_m^*}_{\mathbb{Z}}(0,N)}\left\{jf_{\mathrm{BE}}(N)+f_{\mathrm{BE}}(k)+(M-j-1)f_{\mathrm{BE}}\left(\frac{Mw^*_m-jN-k-1}{M-j-1}\right)\right\}\label{eqn:aliveproconverse-1}\\
&\triangleq&\bar{P}(M,N,w_m^*)\label{eqn:aliveproconverse-2}
\end{IEEEeqnarray}
where $M=\prod_{i=1}^m M^{(i)}$ and $\sum_{i=1}^m n^{(i)}=N$.
\end{theorem}
\begin{IEEEproof}
From \eqref{eqn:berryesseenbound},
the alive probability \eqref{eqn:alivepro} can be upper-bounded as
\begin{equation}\label{eqn:thm1-1}
\Pr\left[\left(\cC^{(i)}\right)_{i=1}^{m}\mbox{ alive}\right]\le\frac{1}{M}\sum_{\bc\in\left(\cC^{(i)}\right)_{i=1}^{m}}{f_{\mathrm{N}}}(\wt(\bc))+\frac{1}{M}\sum_{\bc\in\left(\cC^{(i)}\right)_{i=1}^{m}}{f_{\mathrm{BE}}}(\wt(\bc)).
\end{equation}
Lemmas \ref{lm:LCRC-normalapprox} and \ref{lm:LCRC-berryesseen} tell us that both $f_{\mathrm{N}}$ and $f_{\mathrm{BE}}$ are non-increasing LCRC functions on $[0,\infty)$ and 
$\wt(\bc)\in\{0,1,\ldots,N\}$ for all $\bc\in\left(\cC^{(i)}\right)_{i=1}^{m}$, inequality \eqref{eqn:col_lcrc_integer} from Lemma \ref{lm:lcrc} is then applied 
to obtain \eqref{eqn:aliveproconverse-1} with a given average constraint $w^*_m$.
\end{IEEEproof}

Clearly, $\bar{P}(M,N,w_m^*)$ in \eqref{eqn:aliveproconverse-2} is non-increasing in $w_m^*$.

\emph{Remark:} Equation \eqref{eqn:col_lcrc_integer} basically separates all $M$ points in the summation into three parts, 
$j$ points at $b$, one point at $k$, and $M-j-1$ points at $\left(\frac{Mx^*-jb-k-1}{M-j-1}\right)$.
Similar to Theorem \ref{thm:aliveproconverse}, 
the alive probability is upper-bounded by a quantity that involves $j$ codewords of Hamming weight $N$, 
one codeword of weight $k$, and $M-j-1$ codewords of weight $\left(\frac{Mw^*_m-jN-k-1}{M-j-1}\right)$.
Based on the numerical results, we observe that the maximum of \eqref{eqn:aliveproconverse-1} is always achieved by a $(j,k)\in\setS^{M,w_m^*}_{\mathbb{Z}}(0,N)$ such that $M\gg j$,
which implies that most codewords are of the same Hamming weight. 
Hence, we conjecture that constant composition codes can approach the maximization \eqref{eqn:aliveproconverse-1} with a small gap.  

\subsection{Dynamic Programming Formulation}\label{sec:dpconverse}
Based on the upper bounds of the log-volume and the alive probability, 
the following optimization formula provides a converse bound for channels that wear out:
\begin{equation}\label{eqn:converseVW}
V^*(N,W,\eta) \leq \max_{\substack{ m\in \mathbb{Z}_+,\\ (N_i, W_i)_{i=1}^m\in\setH_{\mathbb{R}}(N,W,m)}}\sum_{i=1}^m\bar{P}\left(M_{\avg}^*(n_i, w_i, \eta), N_i, W_i\right)\log M_{\avg}^*(n_i, w_i, \eta),
\end{equation}
where 
\begin{multline}
\setH_{\mathbb{R}}(N,W,m)=\bigg\{(N_i,W_i)_{i=1}^m: n_i\in \mathbb{Z}_+,\ w_i\in \mathbb{R}_+\mbox{ s.t. }N_i \le N_{i+1},\ W_i \le W_{i+1}\\
\mbox{for all } 1<i<m-1 \mbox{ and } N_m=N,\ W_m=W\bigg\},
\end{multline}
$n_i=N_i-N_{i-1}$, $w_i=W_i-W_{i-1}$, and $N_0=W_0=0$. 
However, the decision variables $(W_i)_{i=1}^m$ cannot be solved for by dynamic programming efficiently,
since $W_i$ is real for all $1\leq i\leq m$. 
To overcome this difficulty, we quantize $w_i$ by segmenting it into several intervals. 
A simple way to perform segmentation is by using the floor operation, i.e, $\lfloor W_i \rfloor\le W_i \le \lfloor W_i \rfloor+1$. 
As mentioned in the previous section, $M_{\avg}^*(n_i, w_i, \eta)$ is a non-decreasing function of $w_i$
and $\bar{P}(M,N_i,W_i)$ is a non-increasing function of $W_i$.
Hence, let $K_i=\lfloor W_i \rfloor$,  we have 
$w_i=W_i-W_{i-1}\le K_{i}-K_{i-1}+1$, which implies that 
\begin{IEEEeqnarray}{rCl}
M_{\avg}^*(n_i,  w_i, \eta)&\leq&M_{\avg}^*(n, K_{i}-K_{i-1}+1, \eta), \\
\bar{P}(M,N_i,W_i) &\le& \bar{P}(M,N_i,K_i).
\end{IEEEeqnarray}
We then conclude that \eqref{eqn:converseVW} can be further upper bounded by 
\begin{multline}
V^*(N,W,\eta) \leq \bar{V}^*(N,W,\eta)\triangleq\max_{\substack{ m\in \mathbb{Z}_+,\\ (N_i, K_i)_{i=1}^m\in\setH_{\mathbb{Z}}(N,W,m)}}\\
\sum_{i=1}^m\bar{P}\left(M_{\avg}^*(n_i, K_i-K_{i-1}+1, \eta), N_i, K_i\right)\log M_{\avg}^*(n_i, K_i-K_{i-1}+1, \eta)\label{eqn:converseVWQ},
\end{multline}
where 
\begin{multline}
\setH_{\mathbb{Z}}(N,W,m)=\bigg\{(N_i,K_i)_{i=1}^m: N_i,K_i\in \mathbb{Z}_+\mbox{ s.t. }N_i \le N_{i+1},\ K_i \le K_{i+1}\\
\mbox{for all } 1<i<m-1 \mbox{ and } N_m=N,\ K_m=W\bigg\}
\end{multline}
and $K_0=0$.
Similar to the dynamic programming procedure for the achievability bound \eqref{eqn:dprec}, we can write \eqref{eqn:converseVWQ} recursively as 
\begin{multline}
\bar{V}^*(N,W,\eta)=\max_{\substack{\text1\leq n\leq N\\0\leq k \leq W}}\bigg\{\bar{V}^*(N-n,W-k,\eta)\\
+\bar{P}\left(M_{\avg}^*(n, k+1, \eta), N, W\right) \log M_{\avg}^*(n, k+1, \eta)\bigg\},
\end{multline}
and 
\begin{equation}\label{eqn:conversev}
\bar{V}^*(N,\eta)=\max_{0< W< N}\bar{V}^*(N,W,\eta).
\end{equation}

Now we have achievable and converse bounds, 
we provide numerical examples to see how close the two bounds are to one another, 
and also gain further insight into the nature of the achievable schemes.

%%%%%%
\section{Numerical  Results}
\label{sec:num}
This section presents some numerical results. 
Here, we consider a BSC with crossover probability $\varepsilon$, denoted by BSC($\varepsilon$),
when the channel is alive, i.e.,
the channel $(\cX, p_\ta, p_\td, \gamma, S, \cY)$ with $\cY=\{0,1,?\}$, $p_\ta(1|0)=p_\ta(0|1)=\varepsilon$,
and $p_\ta(0|0)=p_\ta(1|1)=1-\varepsilon$.
To evaluate $M_\ccc^{*}(n, \wt(P), \eta)$ for BSC($\varepsilon$), 
the normal approximation of \cite[Eq.~(21)]{Moulin2012} (ignoring the $o(1)$ term) is used, i.e.,
\begin{equation}
\log M_\ccc^{*}(n, \wt(P), \eta)\approx nI(P;p_\ta)+\sqrt{n\rho(P;p_\ta)}Q^{-1} (\eta)+
\tfrac{1}{2}\log n + A_\eta(P;p_\ta)+\Delta_\ccc(P;p_\ta),
\end{equation}
where $P$ is the type of input, $I(P;p_\ta)$ is the mutual information, 
$\rho(P;p_\ta)$ is the conditional information variance, $Q^{-1} (\eta)$ is the inverse $Q$-function,
and $A_\eta(P;p_\ta)+\Delta_\ccc(P;p_\ta)$ is the constant part of the approximation in \cite{Moulin2012}.
The $M_\avg^{*}(n, w^*, \eta)$ is evaluated by \cite[Eq.~(1)]{KostinaV2015} (ignoring the $O(1)$ term), i.e.,
\begin{equation}
\log M_{\avg}^*(n, w^*, \eta)\approx nC(w^*/n)-\sqrt{nV(w^*/n)}Q^{-1}(\eta)+\frac{1}{2}\log n,
\end{equation}
where $C(\cdot)$ is the capacity-cost function and $V(\cdot)$ is the dispersion-cost function.\footnote{
For $M_\ccc^{*}(n, \wt(P), \eta)$ and $M_\avg^{*}(n, w, \eta)$, there exist 
achievability \cite[Section III]{Moulin2012}  and converse \cite[Theorem 3]{KostinaV2015}  bounds respectively.
However, computation of these bounds is time-consuming,
especially within dynamic programming.
For ease of numerical computation, the normal approximation is adopted.} 

\begin{figure}[t]
    \centering
    \includegraphics[width=4.5in]{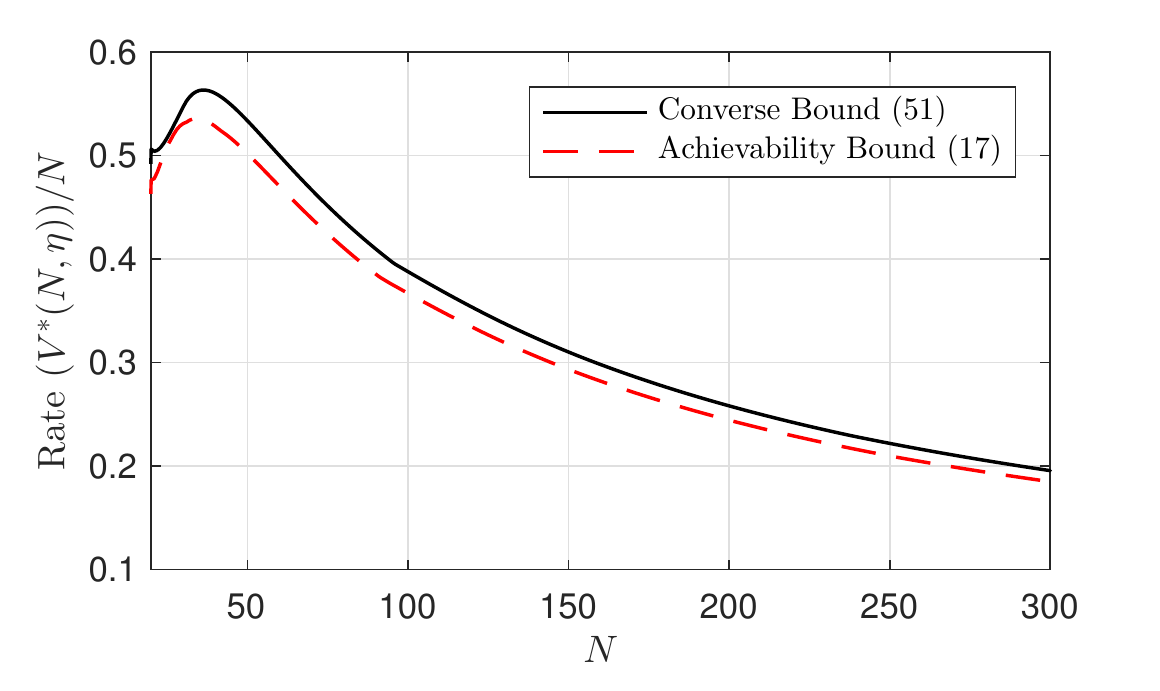}
    \caption{The maximum expected rates for BSC($\varepsilon$).}
    \label{fig:rate}
\end{figure}

The channel is damaged with a probability $\gamma=0.5$ when a $1$ is transmitted, 
and worn out after the amount of damage exceeds $S=5$.
Both the achievable rate \eqref{eqn:volccc_V} and the converse rate \eqref{eqn:conversev} up to $N=400$ 
for a BSC($0.11$) are depicted in Fig.~\ref{fig:rate},
in which the average transmission error was assumed to be lower than $\eta=0.001$.
As expected, the upper bound of alive probability based on the bounded LCRC inequality
implies the upper bound is close to that for a constant composition code.
Hence the converse bound is close to the achievability bound. 
In line with the discussion after Theorem \ref{thm:aliveproconverse}, 
this observation suggests that constant composition codes may achieve the fundamental
communication limits asymptotically. 

\begin{figure}[t]
    \centering
    \includegraphics[width=4.5in]{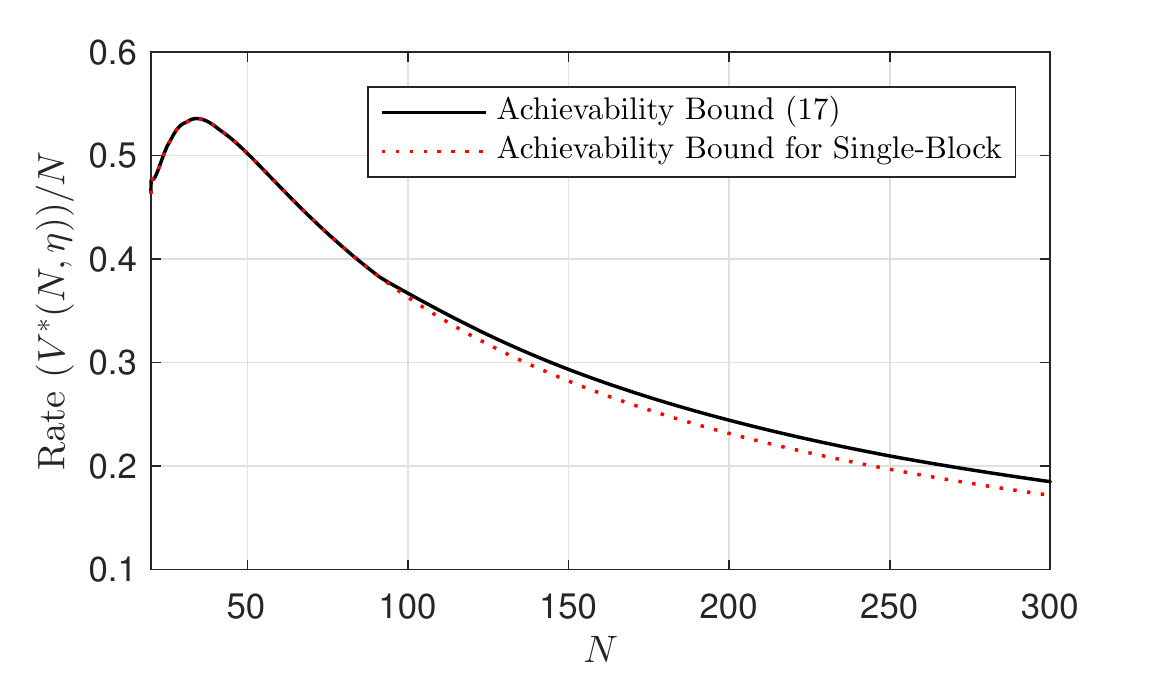}
    \caption{The achievability bound and the achievability bound for single-block transmission.}
    \label{fig:achievabler}
\end{figure}

\begin{figure}[t]
    \centering
    \includegraphics[width=4.5in]{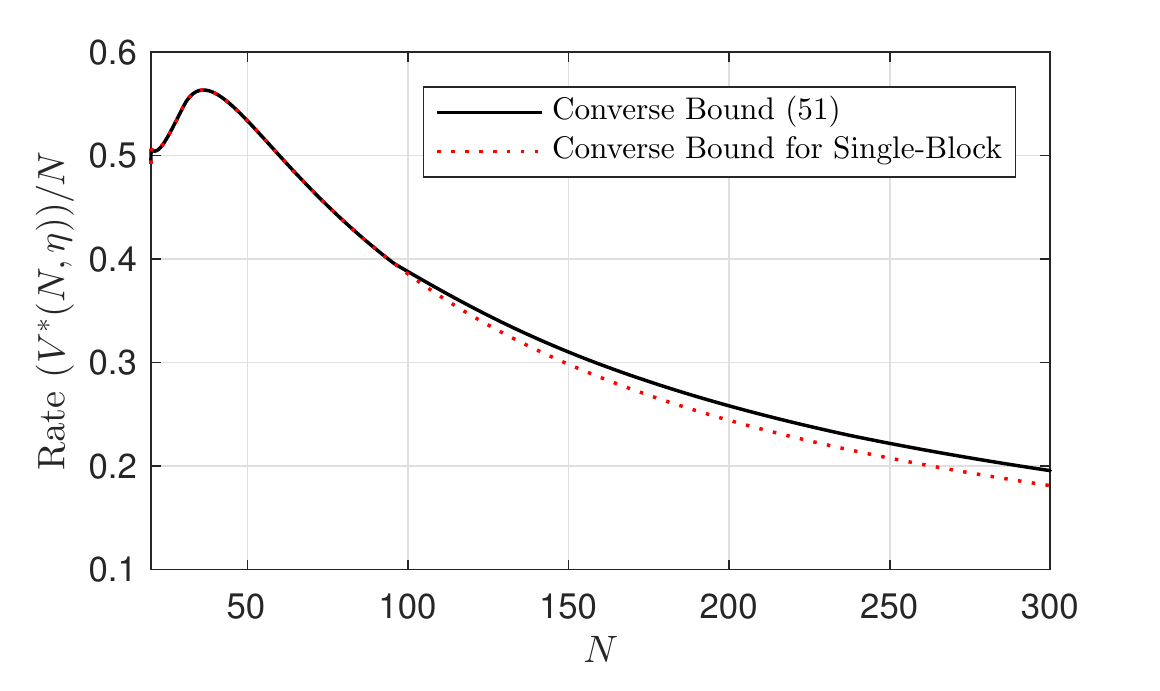}
    \caption{The converse bound and the converse bound for single-block transmission.}
    \label{fig:converser}
\end{figure}

Now we compare multiple-blocks codes with single-block codes using the achievability \eqref{eqn:volccc_V} and converse bounds \eqref{eqn:conversev} over this channel that wears out. 
Fig.~\ref{fig:achievabler} and Fig.~\ref{fig:converser} show the single-block scheme 
performs less well than the multiple-block scheme and the 
gap becomes significant when $N$ increases. 
With the multiple-blocks scheme, segmenting the information can extend the lifetime of channel 
to further increase communication limits.

\begin{figure}[t]
    \centering
    \includegraphics[width=4.5in]{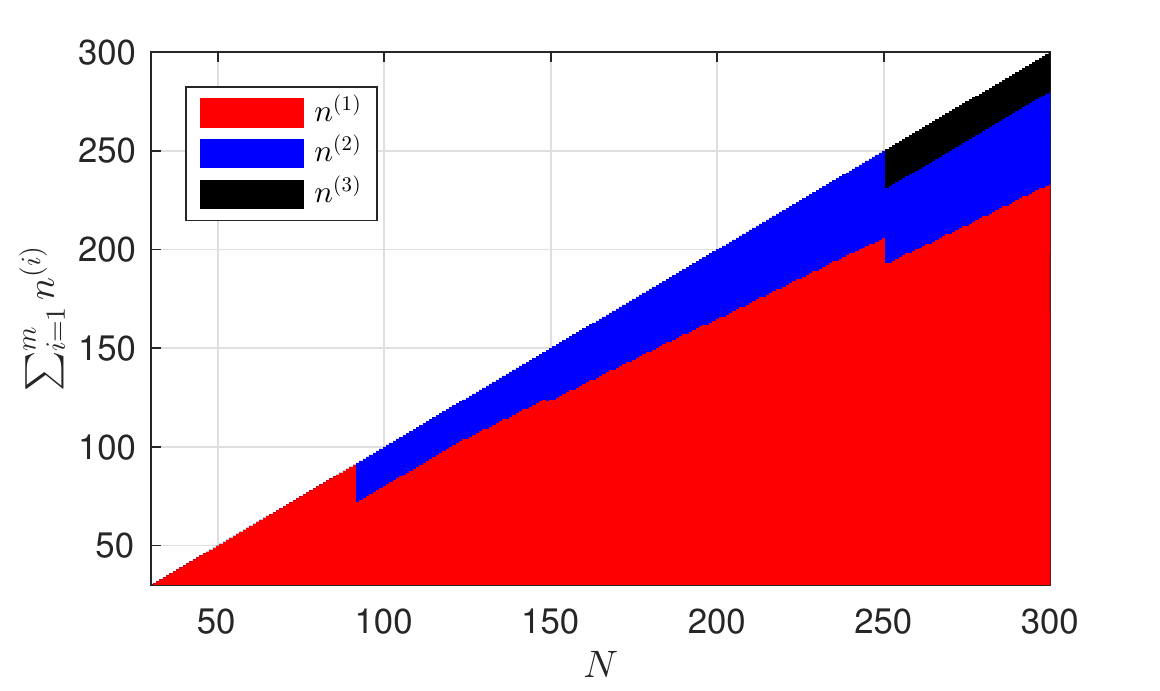}
    \caption{Length of each block for achievability bound.}
    \label{fig:achievablel}
\end{figure}

\begin{figure}[t]
    \centering
    \includegraphics[width=4.5in]{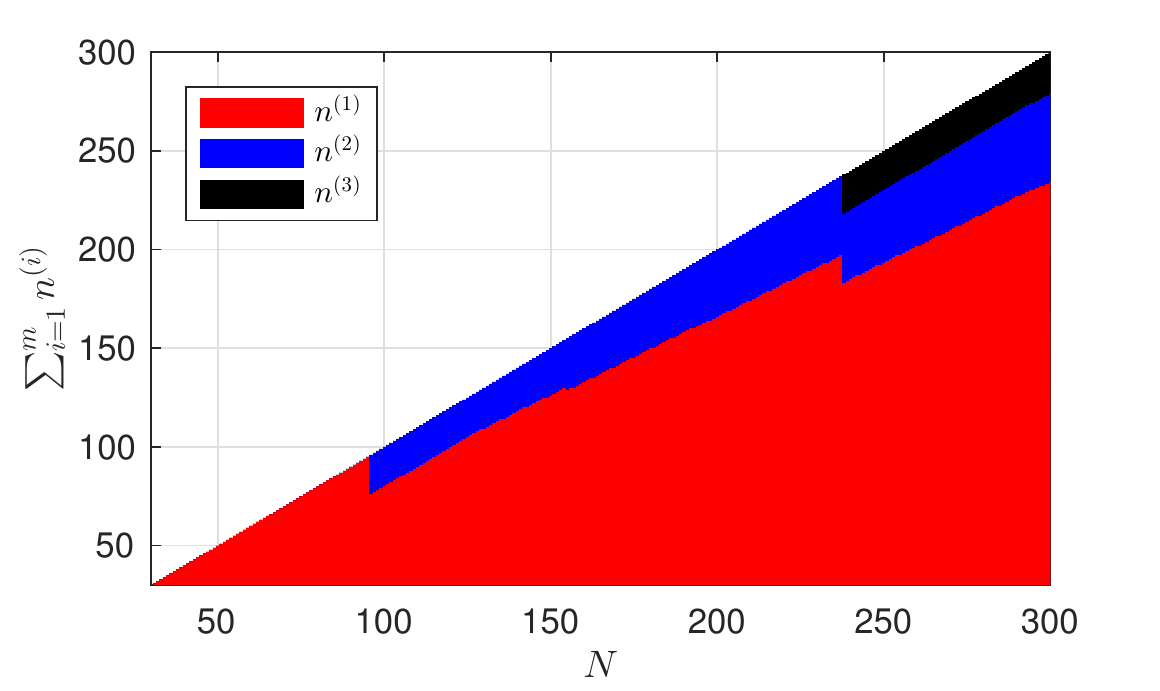}
    \caption{Length of each block for converse bound.}
    \label{fig:conversel}
\end{figure}

\begin{figure}[t]
    \centering
    \includegraphics[width=4.5in]{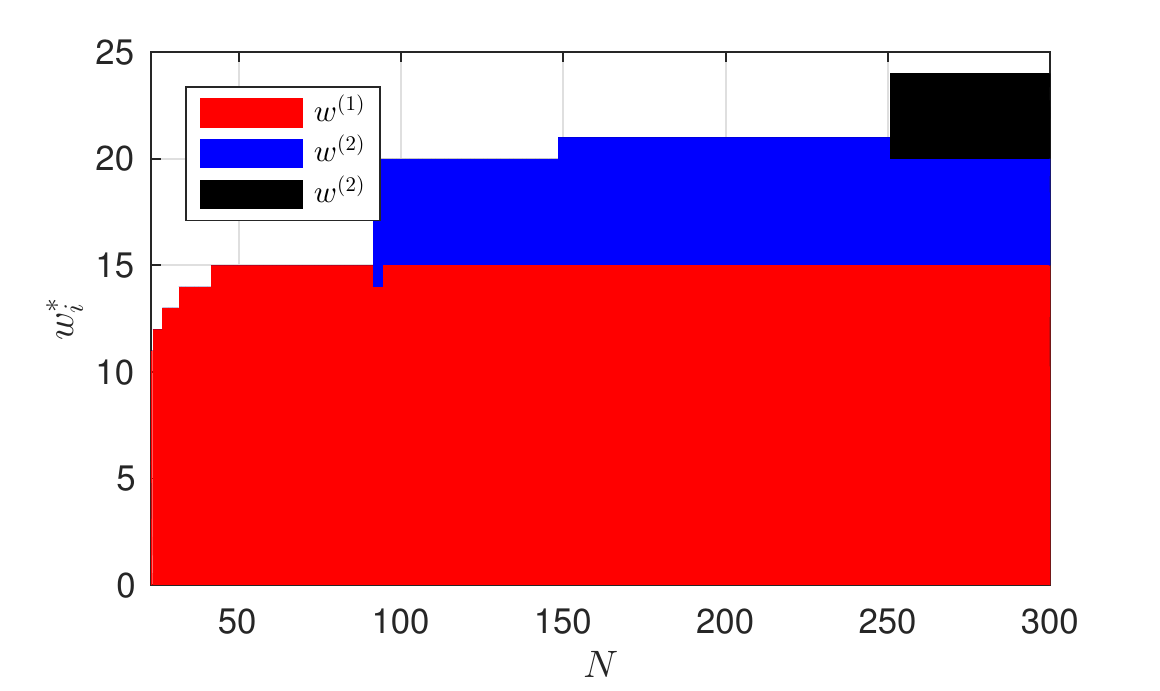}
    \caption{Hamming weight of each block for achievability bound.}
    \label{fig:achievablew}
\end{figure}

\begin{figure}[t]
    \centering
    \includegraphics[width=4.5in]{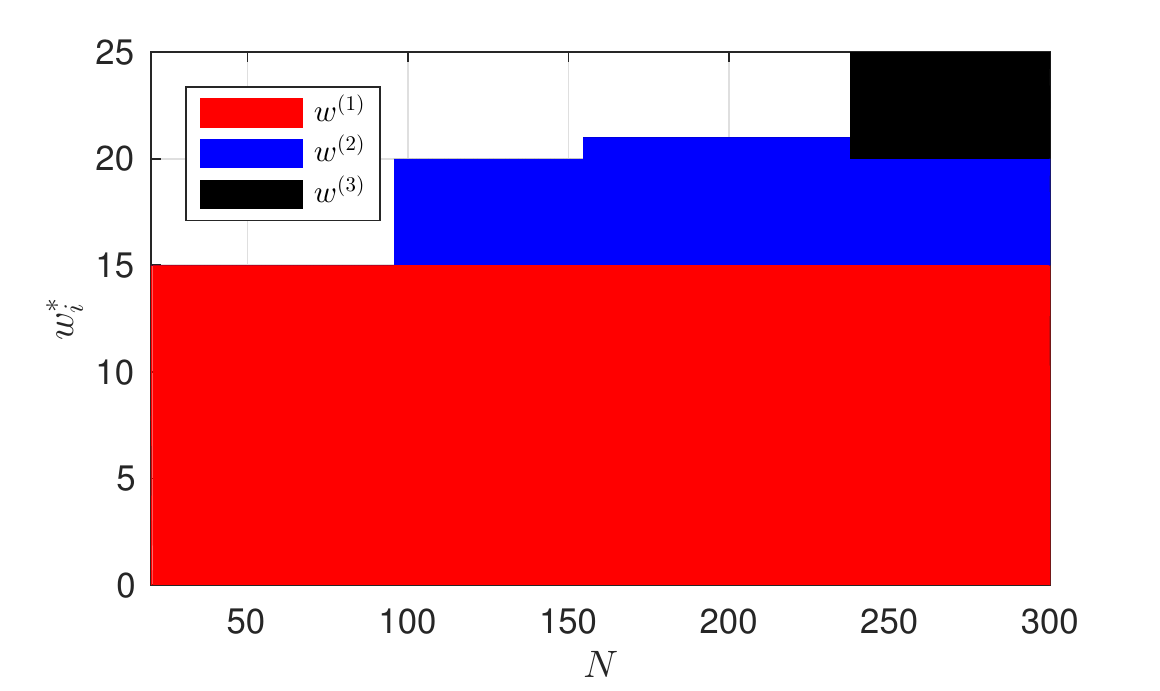}
    \caption{Hamming weight of each block for converse bound.}
    \label{fig:conversew}
\end{figure}

The size of each block based on the achievability and the converse bounds are given in Fig.~\ref{fig:achievablel} and Fig.~\ref{fig:conversel}, respectively. 
In both figures, the length of each transmitted block is plotted in a different color.
As these figures show, when $N$ is small, the single-block scheme is sufficient to optimally transmit information over channels that wear out; 
this is further evidenced by the fact that the curves in Fig.~\ref{fig:achievabler} and \ref{fig:converser} overlap when $N$ is small. 
As $N$ increases, the best strategy is to separate information into blocks;
for example, when $N=300$, both the achievability bound in Fig.~\ref{fig:achievablel} and the converse bound in Fig.~\ref{fig:conversel}
suggest separating information into three blocks of lengths $n^{(1)}$, $n^{(2)}$, and $n^{(3)}$.
The corresponding Hamming weights of the transmitted blocks are given in Fig.~\ref{fig:achievablew} and Fig.~\ref{fig:conversew}.

From Fig.~\ref{fig:achievablel} to \ref{fig:conversew}, we also observe that the lengths and the Hamming weights of the transmitted blocks are non-increasing, i.e.,
$n^{(1)}\ge n^{(2)}\ge \cdots \ge n^{(m)}$ and $w^{(1)}\ge w^{(2)}\ge \cdots \ge w^{(m)}$.
Such an observation is intuitive since the shorter or lighter (Hamming weight) codes are preferred when the channel is about to burn out. 
It remains to determine whether this property holds in general.

%%%%%%%%%%%%%%%%%%%%%%%%
\section{Conclusion and Future Work}\label{sec:con}
Further increasing the connections between reliability theory and information theory, 
we have proposed a model of a channel that wears out and found the maximum expected transmission volume 
that can be achieved using constant composition codes at a given level of average error probability. 
By comparing our achievability result to a novel converse bound, we see that constant composition codes achieve near-optimal performance. 
Dynamic programming formulations are given for computing achievability and converse bounds, and damage state feedback is shown not to 
improve the probability of successive transmission or volume of bits for using constant composition codes. 

An avenue for future work is to consider a channel model where both noisiness and failure probability increase with damage.  
This may model electronic devices that become noisier before they burn out.
%%%%%%%%%%%%%%%%%%%
\appendices
%%%%%%%%%
\section{Proof of Lemma \ref{lm:LCRC-normalapprox}}
\label{append:LCRC-normalapprox}
\begin{figure}[t]
    \centering
    \includegraphics[width=4.5in]{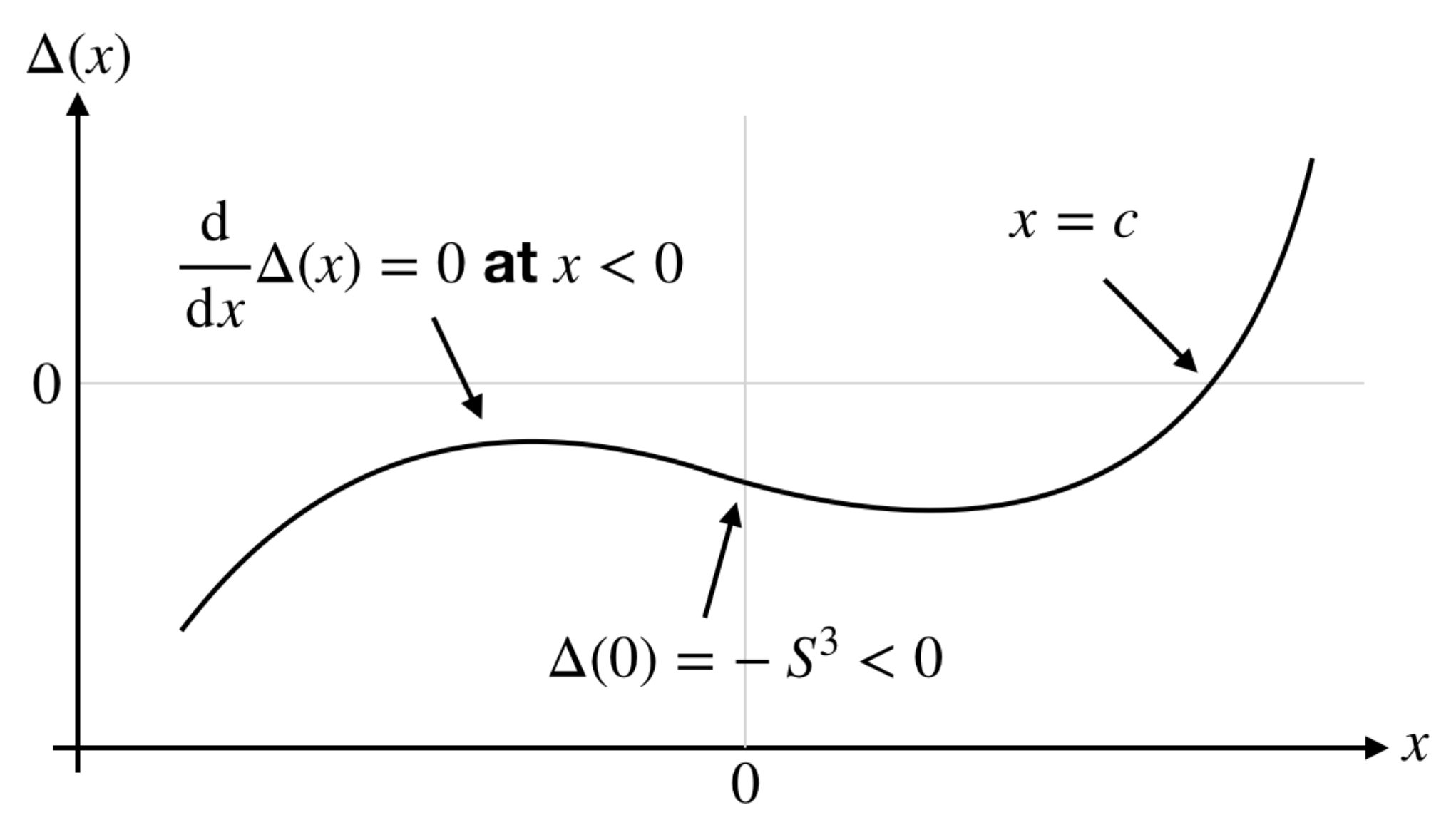}
    \caption{An example of $\Delta(x)$ in \eqref{eqn:secd_approxB}.}
    \label{fig:ex_Delta}
\end{figure}

We first prove the convexity at $x=0$. 
At point $x=0$, $f_\mathrm{N}(x)$ is convex
due to $\lim_{x\downarrow 0}f_\mathrm{N}(x)=1=f_\mathrm{N}(0)$ and $f_\mathrm{N}(x)<f_\mathrm{N}(0)$ for $x>0$.
For $x\in(0,\infty)$, we consider the second derivative of $f_\mathrm{N}(x)$ on $x$.
The second derivative of $f_\mathrm{N}(x)$ for $x\in(0,\infty)$ is 
\begin{equation}\label{eqn:secd_approxB}
\frac{\mathrm{d}^2}{\mathrm{d} x^2} f_\mathrm{N}(x) = \frac{e^{-f(x)^2}}{\sqrt{32x^7\gamma^3(1-\gamma)^3\pi}}\Delta(x),
\end{equation}
where 
\begin{equation}
f(x)=\frac{S-x\gamma}{\sqrt{2x\gamma(1-\gamma)}}
\end{equation}
and
\begin{equation}
\Delta(x)=\gamma^3x^3+(1-\gamma+S)\gamma^2x^2+(3-3\gamma-S)S\gamma x-S^3.
\end{equation}
Since $\Delta(0)=-S^3<0$ and $\frac{\mathrm{d}}{\mathrm{d}x}\Delta(x)=0$ at 
\begin{equation}
x=\frac{-2(1-\gamma+S)\gamma^2\pm\sqrt{4(1-\gamma+S)^2\gamma^4-12(3-3\gamma-S)S\gamma^4}}{6\gamma^3},
\end{equation}
$\frac{\mathrm{d}}{\mathrm{d}x}\Delta(x)=0$ at some negative $x$ as Fig.~\ref{fig:ex_Delta} shows. 
Hence, there must exist a $c>0$ such that $\frac{\mathrm{d}^2}{\mathrm{d} x^2} f_\mathrm{N}(x)\leq 0$ when $0\leq x\leq c$ and  
$\frac{\mathrm{d}^2}{\mathrm{d} x^2} f_\mathrm{N}(x)\geq 0$ when $c\leq x$.
Furthermore, it is clear that $f_\mathrm{N}(x)$ is non-increasing for $x\ge 0$,
which concludes that  $f_\mathrm{N}(x)$ is a non-increasing LCRC function as defined in Def.~\ref{def:LCRC}.

%%%%%%%%%%%
\section{Proof of Lemma \ref{lm:LCRC-berryesseen}}
\label{append:LCRC-berryesseen}
For $x\in [0, x_\mathrm{BE}]$, where $x_\mathrm{BE}\triangleq\max\{x| B_\mathrm{BE}(x)\ge 1\}$ as the definition of $w_\mathrm{BE}$, 
$f_\mathrm{BE}(x)=1$ hence it can be both concave and convex in $[0,x_\mathrm{BE}]$. 
For $x\in [x_\mathrm{BE}, \infty)$, $\frac{\mathrm{d}^2}{\mathrm{d}x^2}f_\mathrm{BE}(x)>0$ hence it is a convex function.
Also, $f_\mathrm{BE}(x)$ is clearly a non-increasing function for $x\ge0$.
Therefore, $f_\mathrm{BE}(x)$ is a non-decreasing LCRC function as defined in Def.~\ref{def:LCRC}.

%%%%%%%%%%%%
\section{Proof of Lemma \ref{lm:karamata}}
\label{append:karamata}
Let 
\begin{equation}
\delta_i\triangleq \frac{f(a_i)-f(b_i)}{a_i-b_i},
\end{equation}
for all $1\leq i \leq M$ and $A_0=B_0=0$.
We have $\delta_i\geq \delta_{i+1}$ due to the convexity of $f$. Then,
\begin{IEEEeqnarray}{rCl}
&&\sum_{i=1}^{M}f(a_i) -f(b_i)\\
&=& \sum_{i=1}^M \delta_i (a_i-b_i)\\
&=& \sum_{i=1}^M \delta_i \big[(A_i-A_{i-1})-(B_i-B_{i-1})\big]\\
&=& \sum_{i=1}^M \delta_i (A_i-B_i)-\sum_{i=1}^M \delta_i(A_{i-1}-B_{i-1}) \\
&=& \delta_M\underbrace{(A_M-B_M)}_{=0}+\sum_{i=1}^{M-1}(\delta_i-\delta_{i+1})(A_i-B_i)-\delta_1\underbrace{(A_0-B_0)}_{=0}\\
&=&\sum_{i=1}^{M-1}\underbrace{(\delta_i-\delta_{i+1})}_{\geq0}\underbrace{(A_i-B_i)}_{\geq 0} \geq 0.
\end{IEEEeqnarray}

%%%%%%%%%%%% Majorsequence
\section{Proof of Lemma \ref{lm:majorsequence}}
\label{append:majorsequence}
Let ${\boldsymbol x}=(x_1,x_2,\ldots,x_M)$ be the sequence in \eqref{eqn:majorsequence}. 
Suppose there exists a sequence ${\boldsymbol y}=(y_1,y_2,\ldots,y_M)\in[c,b]^M$
such that $y_1\ge y_2 \cdots \ge y_M$, $\sum_{i=1}^M y_i=Mx^*$, and ${\boldsymbol y}$ majorizes ${\boldsymbol x}$, i.e.,
\begin{equation}\label{eqn:major}
\sum_{i=1}^k y_i \ge \sum_{i=1}^k x_i
\end{equation}
for all $1\le k\le M-1$.
Clearly $y_i=x_i=b$ for all $i\le j$ since all $x_i, y_i\in[c,b]$.
Now consider the case of $y_{j+1}>x_{j+1}$, which means that $\sum_{i=j+2}^M y_i < (M-j-1)c$
and there must be a $y_i$ for $i\ge j+2$ such that $y_i< c$. 
Therefore, $y_{j+1}$ must equal $x_{j+1}$ to satisfy \eqref{eqn:major} for all $k\le j+1$. 
For all $i\ge j+2$, equalities $y_{i}=x_i$ can be verified in the same way. 
Hence, \eqref{eqn:major} holds for all $k$ if and only if $y_i=x_i$ for all $i$,
which proves that the sequence \eqref{eqn:majorsequence} majorizes all sequences from $[c,b]^M$ with the average $x^*$.

%%%%%%%%%%% LCRC
\section{Proof of Lemma \ref{lm:lcrc}}
\label{append:lcrc}
Without loss of generality, we assume $x_i\ge x_{i+1}$ for all $1\leq i < M$.
Suppose $c\geq x_1$, which means that all $x_i$ are located in the concave region of $f$. 
Therefore Jensen's inequality can be applied to upper bound $\sum_{i=1}^M f(x_i)$ as
\begin{equation}
\sum_{i=1}^M f(x_i)\leq Mf(x^*)=f(x^*)+(M-1)f\bigg(\frac{Mx^*-x^*}{M-1}\bigg)
\end{equation}
which is a special case of \eqref{eqn:col_lcrc_real} when $j=0$ and $r=x^*$.

We then consider the case of $x_1>c$.
Let $t$ be the index such that 
\begin{equation}\label{eqn:lcrc-1}
x_1\ge \cdots \ge x_t \ge c > x_{t+1}\ge \cdots \ge x_M,
\end{equation}
and $\sum_{i=1}^t x_i=s$.   
Since all $x_1 , x_2, \ldots, x_t$ are located in the convex region $[c,b]$ of the function $f$, 
from Lemmas \ref{lm:karamata} and \ref{lm:majorsequence} we have
\begin{equation}\label{eqn:lcrc-2}
\sum_{i=1}^t f(x_i)\leq jf(b)+f(r)+(t-j-1)f(c),
\end{equation} 
where $j=\Big\lfloor \frac{s-tc}{b-c}\Big\rfloor$ and $r=s-jb-(t-j-1)c$.
Then we upper bound the summation of all points located in concave region $[a,c]$ by Jensen's inequality as
\begin{equation}\label{}
(t-j-1) f(c) +\sum_{i=t+1}^{M}f(x_i)
\leq (M-j-1) f\Bigg(\frac{(t-j-1)c+\sum_{i=t+1}^{M}x_i}{M-j-1}\Bigg).\label{eqn:lcrc-3}
\end{equation}
Combing both inequalities \eqref{eqn:lcrc-2} and \eqref{eqn:lcrc-3}, we have
\begin{IEEEeqnarray}{rCl}
\sum_{i=1}^M f(x_i) &\leq& jf(b)+f(r)+(M-j-1) f\Bigg(\frac{(t-j-1)c+\sum_{i=t+1}^{M}x_i}{M-j-1}\Bigg)\label{eqn:lcrc-4}\\
&=& jf(b)+f(r)+(M-j-1) f\Bigg(\frac{Mx^*-jb-r}{M-j-1}\Bigg)\label{eqn:lcrc-5},
\end{IEEEeqnarray}
for some $j$ and $r$ such that 
\begin{equation}\label{eqn:lcrc-6}
b\ge r \ge \frac{Mx^*-jb-r}{M-j-1}\ge a.
\end{equation}
By taking the maximum of \eqref{eqn:lcrc-5} over all possible $j$ and $r$ that satisfy \eqref{eqn:lcrc-6} yields \eqref{eqn:col_lcrc_real}.

Now we consider the case when $f$ is also non-increasing.  
Let $k=\lfloor r\rfloor$,
$f(k)\ge f(r)$ and $f\left(\frac{Mx^*-jb-k-1}{M-j-1}\right) \ge f\left(\frac{Mx^*-jb-r}{M-j-1}\right)$ due to $k+1\ge r \ge k$.
Moreover, constraint \eqref{eqn:lcrc-6} can be rewritten as 
\begin{equation}
k+1\ge r \ge \frac{Mx^*-jb-r}{M-j-1} \ge \max\left\{\frac{Mx^*-jb-k-1}{M-j-1},a\right\}.
\end{equation}
Therefore, $(j,r)\in \setS^{M,x^*}_{\mathbb{R}}(a,b)$ implies $(j, \lfloor r\rfloor)\in \setS^{M,x^*}_{\mathbb{Z}}(a,b)$.
However, constraint \eqref{eqn:lcrc-6} may result in an $r$ such that $\lfloor r\rfloor <a$, which makes $f(k)$ undefined. 
To avoid that, we restrict $a$ to be an integer and this establishes \eqref{eqn:col_lcrc_integer}.

\bibliographystyle{IEEEtran} 
\bibliography{abrv,conf_abrv,lrv_lib}

\newcommand{\SortNoop}[1]{}
\begin{thebibliography}{10}
\providecommand{\url}[1]{#1}
\csname url@samestyle\endcsname
\providecommand{\newblock}{\relax}
\providecommand{\bibinfo}[2]{#2}
\providecommand{\BIBentrySTDinterwordspacing}{\spaceskip=0pt\relax}
\providecommand{\BIBentryALTinterwordstretchfactor}{4}
\providecommand{\BIBentryALTinterwordspacing}{\spaceskip=\fontdimen2\font plus
\BIBentryALTinterwordstretchfactor\fontdimen3\font minus
  \fontdimen4\font\relax}
\providecommand{\BIBforeignlanguage}[2]{{%
\expandafter\ifx\csname l@#1\endcsname\relax
\typeout{** WARNING: IEEEtran.bst: No hyphenation pattern has been}%
\typeout{** loaded for the language `#1'. Using the pattern for}%
\typeout{** the default language instead.}%
\else
\language=\csname l@#1\endcsname
\fi
#2}}
\providecommand{\BIBdecl}{\relax}
\BIBdecl

\bibitem{WuVT2017}
T.-Y. Wu, L.~R. Varshney, and V.~Y.~F. Tan, ``Communication over a channel that
  wears out,'' in \emph{Proc. 2017 IEEE Int. Symp. Inf. Theory}, Jun. 2017, pp.
  581--585.

\bibitem{Nakagawa2007}
T.~Nakagawa, \emph{Shock and Damage Models in Reliability Theory}.\hskip 1em
  plus 0.5em minus 0.4em\relax London: Springer-Verlag, 2007.

\bibitem{NairD2015}
G.~B. Nair and S.~J. Dhoble, ``A perspective perception on the applications of
  light-emitting diodes,'' \emph{Luminescence}, vol.~30, no.~8, pp. 1167--1175,
  Dec. 2015.

\bibitem{VarshneyMG2012}
L.~R. Varshney, S.~K. Mitter, and V.~K. Goyal, ``An information-theoretic
  characterization of channels that die,'' \emph{{IEEE} Trans. Inf. Theory},
  vol.~58, no.~9, pp. 5711--5724, Sep. 2012.

\bibitem{PursleyS1989}
M.~B. Pursley and S.~D. Sandberg, ``Variable-rate coding for meteor-burst
  communications,'' \emph{{IEEE} Trans. Commun.}, vol.~37, no.~11, pp.
  1105--1112, Nov. 1989.

\bibitem{Ryan1997}
W.~E. Ryan, ``Optimal signaling for meteor burst channels,'' \emph{{IEEE}
  Trans. Commun.}, vol.~45, no.~5, pp. 489--496, May 1997.

\bibitem{KochLS2009}
T.~Koch, A.~Lapidoth, and P.~P. Sotiriadis, ``Channels that heat up,''
  \emph{{IEEE} Trans. Inf. Theory}, vol.~55, no.~8, pp. 3594--3612, Aug. 2009.

\bibitem{OzelUG2016}
O.~Ozel, S.~Ulukus, and P.~Grover, ``Energy harvesting transmitters that heat
  up: Throughput maximization under temperature constraints,'' \emph{{IEEE}
  Trans. Wireless Commun.}, vol.~15, no.~8, pp. 5440--5452, Aug. 2016.

\bibitem{Gallager1968}
R.~G. Gallager, \emph{Information Theory and Reliable Communication}.\hskip 1em
  plus 0.5em minus 0.4em\relax New York: John Wiley \& Sons, 1968.

\bibitem{Gallager1972}
R.~Gallager, \emph{Information Theory and Reliable Communication}, ser.
  International Centre for Mechanical Sciences, Courses and Lectures.\hskip 1em
  plus 0.5em minus 0.4em\relax Vienna: Springer-Verlag, 1972, no.~30.

\bibitem{Weiss1960}
L.~Weiss, ``On the strong converse of the coding theorem for symmetric channels
  without memory,'' \emph{Q. Appl. Math.}, vol.~18, no.~3, pp. 209--214, Oct.
  1960.

\bibitem{Strassen1962}
V.~Strassen, ``Asymptotische absch{\"{a}}tzungen in {S}hannon's
  informationstheorie,'' in \emph{Transactions of the 3rd Prague Conference on
  Information Theory, Statistical Decision Functions, Random Processes}.\hskip
  1em plus 0.5em minus 0.4em\relax Prague: Pub. House of the Czechoslovak
  Academy of Sciences, 1962, pp. 689--723.

\bibitem{PolyanskiyPV2010}
Y.~Polyanskiy, H.~V. Poor, and S.~{Verd\'{u}}, ``Channel coding rate in the
  finite blocklength regime,'' \emph{{IEEE} Trans. Inf. Theory}, vol.~56,
  no.~5, pp. 2307--2359, May 2010.

\bibitem{TomamichelT2013}
M.~Tomamichel and V.~Y.~F. Tan, ``A tight upper bound for the third-order
  asymptotics for most discrete memoryless channels,'' \emph{{IEEE} Trans. Inf.
  Theory}, vol.~59, no.~11, pp. 7041--7051, Nov. 2013.

\bibitem{Moulin2017}
P.~Moulin, ``The log-volume of optimal codes for memoryless channels,
  asymptotically within a few nats,'' \emph{{IEEE} Trans. Inf. Theory},
  vol.~63, no.~4, pp. 2278--2313, Apr. 2017.

\bibitem{Fano1961}
R.~M. Fano, \emph{Transmission of Information: A Statistical Theory of
  Communications}.\hskip 1em plus 0.5em minus 0.4em\relax Cambridge, MA: MIT
  Press, 1961.

\bibitem{KostinaV2015}
V.~Kostina and S.~{Verd\'{u}}, ``Channels with cost constraints: Strong
  converse and dispersion,'' \emph{{IEEE} Trans. Inf. Theory}, vol.~61, no.~5,
  pp. 2415--2429, May 2015.

\bibitem{Moulin2012}
P.~Moulin, ``The log-volume of optimal constant-composition codes for
  memoryless channels, within $o(1)$ bits,'' in \emph{Proc. 2012 IEEE Int.
  Symp. Inf. Theory}, Jul. 2012, pp. 826--830.

\bibitem{ScarlettMF2015}
J.~Scarlett, A.~Martinez, and A.~{{Guill\'{e}n i F\`{a}bregas}}, ``Refinements
  of the third-order term in the fixed error asymptotics of
  constant-composition codes,'' in \emph{Proc. 2015 IEEE Int. Symp. Inf.
  Theory}, Jun. 2015, pp. 2954--2958.

\bibitem{TandonMV2016}
A.~Tandon, M.~Motani, and L.~R. Varshney, ``Subblock-constrained codes for
  real-time simultaneous energy and information transfer,'' \emph{{IEEE} Trans.
  Inf. Theory}, vol.~62, no.~7, pp. 4212--4227, Jul. 2016.

\bibitem{AltugW2014}
Y.~Alt{\u{u}}g and A.~B. Wagner, ``Feedback can improve the second-order coding
  performance in discrete memoryless channels,'' in \emph{Proc. 2014 IEEE Int.
  Symp. Inf. Theory}, Jul. 2014, pp. 2361--2365.

\bibitem{PolyanskiyW}
\BIBentryALTinterwordspacing
Y.~Polyanskiy and Y.~Wu, ``Lecture notes on information theory.'' [Online].
  Available: \url{{http://people.lids.mit.edu/yp/homepage/papers.html}}
\BIBentrySTDinterwordspacing

\bibitem{Schulz2016}
J.~Schulz, ``The optimal {B}erry--{E}sseen constant in the binomial case,''
  Ph.D. dissertation, Universit{\"a}t Trier, Germany, Jun. 2016.

\bibitem{Cirtoaje2006}
V.~C\^{i}rtoaje, \emph{Algebraic Inequalities: Old and New Methods}.\hskip 1em
  plus 0.5em minus 0.4em\relax GIL Publishing House, 2006.

\bibitem{Karamata1932}
J.~Karamata, ``Sur une in{\'e}galit{\'e} relative aux fonctions convexes,''
  \emph{Publications de l'Institut Math{\'e}matique}, vol.~1, no.~1, pp.
  145--147, 1932.

\end{thebibliography}

\end{document}